\documentclass[reqno]{amsart}
\usepackage{amscd,amssymb,url}

\DeclareMathOperator{\End}{End}
\DeclareMathOperator{\tr}{tr}
\DeclareMathOperator{\Mod}{Mod}
\DeclareMathOperator{\Fred}{Fred}
\DeclareMathOperator{\Ch}{Ch}

\DeclareMathOperator{\Tr}{Tr}
\DeclareMathOperator{\Pro}{Pro}
\DeclareMathOperator{\Lin}{Lin}
\DeclareMathOperator{\Bun}{Bun}
\DeclareMathOperator{\ind}{ind}

\theoremstyle{plain}
\newtheorem{theorem}{Theorem}[section]
\newtheorem{lemma}[theorem]{Lemma}
\newtheorem{proposition}[theorem]{Proposition}

\theoremstyle{definition}
\newtheorem{definition}[theorem]{Definition}

\theoremstyle{remark}

\newtheorem{note}{Note}[section]

\newtheorem{remark}{Remark}[section]

\numberwithin{equation}{section}
\numberwithin{figure}{section}

\newcommand{\cH}{{\mathcal H}}

\newcommand{\cE}{{\mathcal E}}
\newcommand{\cK}{{\mathcal K}}

\newcommand{\cP}{{\mathcal P}}

\newcommand{\CC}{{\mathbb C}}
\newcommand{\RR}{{\mathbb R}}
\newcommand{\ZZ}{{\mathbb Z}}
\newcommand{\QQ}{{\mathbb Q}}

\newcommand{\redK}{\widetilde K} % reduced K theory

\newcommand{\RP}{\RR {\rm P}}

\newcommand{\UK}{U{_\mathcal K}}

\newcommand{\ad}{\text{ad}}

\begin{document}

\title[Twisted K-theory and K-theory of bundle gerbes]{Twisted K-theory
and K-theory of bundle gerbes}

\author[P. Bouwknegt]{Peter Bouwknegt}
\address[Peter Bouwknegt]
{Department of Physics and Mathematical Physics
and Department of Pure Mathematics\\
University of Adelaide\\
Adelaide, SA 5005 \\
Australia}
\email{pbouwkne@physics.adelaide.edu.au, pbouwkne@maths.adelaide.edu.au}

\author[A.L. Carey]{Alan L. Carey}
\address[Alan L. Carey]
{Department of Pure Mathematics\\
University of Adelaide\\
Adelaide, SA 5005 \\
Australia}
\email{acarey@maths.adelaide.edu.au}

\author[V. Mathai]{Varghese Mathai}
\address[Varghese Mathai]
{Department of Mathematics\\
MIT \\
Cambridge, MA 02139 \\
USA and Department of Pure Mathematics\\
University of Adelaide\\
Adelaide, SA 5005 \\
Australia}
\email{vmathai@math.mit.edu, vmathai@maths.adelaide.edu.au }

\author[M.K. Murray]{Michael K. Murray}
\address[Michael K. Murray]
{Department of Pure Mathematics\\
University of Adelaide\\
Adelaide, SA 5005 \\
Australia}
\email{mmurray@maths.adelaide.edu.au}

\author[D. Stevenson]{Danny Stevenson}
\address[Danny Stevenson]
{Department of Pure Mathematics\\
University of Adelaide\\
Adelaide, SA 5005 \\
Australia}
\email{dstevens@maths.adelaide.edu.au}

\thanks{The authors acknowledge the support of the Australian
Research Council.  In addition PB acknowledges support from the
Caltech/USC Center for Theoretical Physics, VM and ALC from the
Clay Mathematics Institute and ALC from the Max Planck Institute,
Albert Einstein, Potsdam.}

\subjclass{81T30, 19K99}

\begin{abstract}
In this note we introduce the notion of bundle gerbe $K$-theory
and investigate the relation to twisted $K$-theory.  We provide
some examples.  Possible applications of bundle gerbe $K$-theory to
the classification of $D$-brane charges in nontrivial backgrounds are
briefly discussed.
\end{abstract}
\maketitle

% ----------------------------------------------------------------------
\section{Introduction}

Based on explicit calculations of $D$-brane charges and the analysis of
brane creation-annihilation processes it has been argued that
$D$-branes, in the absence of background $B$-fields, carry charges which
take values in $K$-theory \cite{MM,Wit1,Hor,MW}.  (For background on
$D$-branes see, e.g., \cite{Pol}.)  This proposal has been extended
to incorporate nontrivial background $B$-fields in \cite{Wit1,Kap}
for torsion $B$-fields, and in \cite{BM,Ati2} for general $B$-fields,
in which case twisted $K$-theory \cite{Ros} is needed.
The picture of $D$-brane charges taking values in (twisted)
$K$-theory has received further support from an analysis of $M$-theory
\cite{DMW}, noncommutative tachyons \cite{Wit2,HM} and explicit
examples (see, e.g., \cite{FS} and references therein).

On the other hand, since $B$-fields are most naturally
described as connections over 1-gerbes,
it has been clear for some time that gerbes are relevant to
understanding the properties of $D$-branes in string theory.
The occurrence of gerbes
can, for instance, be inferred from the anomaly cancellation
argument in \cite{FW} and is mentioned explicitly in \cite{HM}.

We believe that gerbes play a role in string theory which is yet to be
fully understood. The aim of this note is to argue that the twisted
$K$-theory of a pair $(M,[H])$, where $M$ is a manifold and $[H]$ is
an integral \v Cech class, can be obtained from the
$K$-theory of a special kind of gerbe over $M$, namely the bundle
gerbes of \cite{Mur}.  In this paper, for the
first time, we introduce the notion of a bundle gerbe
module, which, in a sense, can also be thought of as a twisted
vector bundle or non-abelian gerbe (see \cite{Kalk} for an
earlier proposal), and define the $K$-theory of bundle gerbes as
the Grothendieck group of the semi-group of bundle gerbe modules.
We show that bundle gerbe $K$-theory is isomorphic
to twisted $K$-theory, whenever $[H]$ is a torsion class in $H^3(M,
\mathbb Z)$.
When $[H]$ is not a torsion class in $H^3(M, \mathbb Z)$ we consider the
lifting bundle gerbe associated to the $PU(\cH)$ bundle with Dixmier-Douady
class $[H]$ and in this case we prove
that twisted $K$-theory
is the Grothendieck group of the semi-group of  $\UK$-bundle gerbe modules,
which are the infinite dimensional cousins of bundle gerbe modules.
   It remains to understand how it might be used in string theory for
example whether the analysis of
\cite{FW} applies in the case where the background $B$-field does not
define a torsion class in
$H^3(M,\ZZ)$ (related issues have recently been discussed in 
\cite{MalMooSei}).

This note is organised as follows. Section 2 summarises the theory of
bundle gerbes. These are geometric objects that are associated with
degree 3 integral \v Cech cohomology classes on $M$. The notion of stable
equivalence of bundle gerbes, which is essential for the understanding
of the sense in which the degree 3 class (known as the Dixmier-Douady
class of the bundle gerbe) determines an associated bundle gerbe is
the subject of Section 3.  The $K$-theory of bundle gerbes is
introduced in Sections 4 and 5 and in Section 6 we analyse
characteristic classes of bundle gerbe modules.   
Twisted $K$-theory in its various
manifestations is described in Section 6 where we prove that the
bundle gerbe $K$-theory is isomorphic to twisted $K$-theory in the torsion
case and analyse characteristic classes of bundle gerbe modules.
In Section 7 we consider bundle gerbes with non-torsion
Dixmier-Douady class  and show that twisted $K$-theory  is isomorphic
to the  $\UK$ bundle gerbe $K$-theory
of the lifting bundle gerbe.
We extend our discussion of characteristic classes for bundle 
gerbe modules to the non-torsion case in Section 9, where 
we also discuss twisted cohomology.
In Section 8 we calculate some examples of twisted $K$-theory,
and we conclude with some remarks in Section 10.

While completing this note a preprint \cite{LupUri} appeared which
uses similar ideas in the context of the $K$-theory of orbifolds and
another \cite{Mac} which introduces twisted vector and principal
bundles which are the same as our bundle gerbe modules when the bundle
gerbe arises from an open cover.

% ----------------------------------------------------------------------
\section{Bundle gerbes}
\label{bundle_gerbes}

\subsection{Bundle gerbes and Dixmier-Douady classes}

Before recalling the definition of bundle gerbe from \cite{Mur} we
need some notation for fibre products.  Mostly we will be working with
smooth manifolds and smooth maps but often these will need to be
infinite-dimensional. In the interest of brevity we will just say map.

We will be interested in maps  $\pi \colon Y \to M$ which admit local
sections.  That is, for every $x \in M$ there is an open set $U  $
containing $x$ and  a local section $s \colon U \to Y$.
          We call such maps locally split.   Note that a locally split map is
necessarily
surjective.  
Locally trivial fibrations are, of course, locally split, but the
converse is not true. Indeed one case of particular
interest will be when $M$ has an open cover $\{U_i \}_{i \in I}$ and
$$
Y = \{ (x, i) \mid x \in U_i \}
$$
the
disjoint union of all the open sets $U_i$ with $\pi(x, i) =x$.
          This example is locally split by $s_i \colon U_i \to Y$, with
$s_i(x) = (x, i)$ but it is rarely a fibration.

Let $\pi \colon Y \to M$ be locally split. Then we denote by
          $Y^{[2]} = Y\times_\pi Y$ the fibre product
of $Y$ with itself over $\pi$,
          that is the subset of pairs $(y, y')$ in
$Y \times Y$ such that $\pi(y) = \pi(y')$.  More generally we denote
the $p$th fold fibre product by $Y^{[p]}$.

Recall that a hermitian line bundle  $L \to M$ is a complex line
bundle with a fibrewise
hermitian inner product. For such a line bundle the set of all
vectors of norm $1$ is
a principal $U(1)$ bundle. Conversely if $P \to M$ is a principal
$U(1)$ bundle then associated to it is a complex
line bundle with fibrewise hermitian inner product.  This is
formed in the standard way as the quotient of $P \times \CC$ by the
action of $U(1)$ given by $(p, z)w = (pw, w^{-1}z)$ where $w \in U(1)$.
       The theory of
bundle gerbes as developed in \cite{Mur} used principal bundles
(actually $\CC^\times$ bundles) but it can be equivalently
expressed in terms of hermitian line bundles. In the
discussion below we will mostly adopt this perspective.
All maps between hermitian line bundles will be assumed
to preserve the inner product unless we explicitly comment otherwise.

A bundle gerbe\footnote{Strictly speaking what we are about to define
should be called a hermitian  bundle gerbe but the extra terminology is
overly burdensome.} over $M$ is a pair $(L, Y)$ where
$\pi \colon Y \to M$ is a locally split map and
$L$ is a hermitian line bundle $L \to Y^{[2]}$ with a product, that is,
a hermitian  isomorphism
$$
L_{(y_1, y_2)} \otimes L_{(y_2, y_3)} \to L_{(y_1, y_3)}
$$
for every $(y_1, y_2)$ and $(y_2, y_3)$ in $Y^{[2]}$.
We require the product to be smooth in $y_1$, $y_2$ and
$y_3$ but in the interests of brevity we will not state the various
definitions needed to make this requirement precise, they  can be found in
\cite{Mur}.
The product is required to be
associative whenever triple products are defined. Also in \cite{Mur}
it is shown that the existence of the product and the associativity
imply isomorphisms $L_{(y, y)} \simeq \CC$ and $L_{(y_1, y_2)} \simeq
L_{(y_2, y_1)}^*$. We shall
often refer to a bundle gerbe $(L, Y)$ as just $L$.

Various operations are possible on bundle gerbes.  Let $(L, Y)$
be a bundle gerbe over $M$.
Let $\pi \colon Z \to N$ be another locally split map
and let  $\hat\phi \colon Z \to Y$ be a fibre map covering
a map $\phi \colon N \to M$.  Then there is an induced map
${\hat \phi}^{[2]} \colon Z^{[2]} \to Y^{[2]}$ which can
          be used to pull-back the bundle $L \to Y^{[2]}$ to a bundle
$({\hat\phi}^{[2]})^{-1}(L) \to Z^{[2]}$.  This has an induced product on it
and defines a bundle gerbe which we denote, for simplicity,
by $(\phi^{-1}(L), Z)$ or $\phi^{-1}(L)$.  Two special cases
of this are important. The first is when we just have a map
          $f\colon N \to M$ and use this to pull-back $Y \to M$ to
$f^{-1}(Y) \to N$. The second is when we have $M = N$ and
$\phi$ the identity.

If $(L, Y)$ is a bundle gerbe  we can define a
new bundle gerbe, $(L^*, Y)$, the dual of $(L, Y)$, by taking
the dual of $L$.
Also if  $(L, Y)$ and $(J, Z)$ are two bundle gerbes we can define their
product $(L\otimes J,  Y\times_\pi Z)$ where $Y\times_\pi Z
=\{ (y, z) \colon \pi_Y(y) = \pi_Z(z) \} $ is
the fibre product of $Y$ and $Z$ over their projection maps.

A morphism from a bundle gerbe  $(L, Y)$ to a bundle
gerbe  $(J, Z)$ consists of a pair of maps $(g, f)$ where
           $f \colon Y \to Z$ is a map commuting with the
projection to $M$ and  $g \colon L \to J$ is a bundle map covering the
induced map $f^{[2]} \colon Y^{[2]} \to Z^{[2]}$ and commuting with the
bundle gerbe products on $J$ and $L$ respectively. If $f$ and $g$ are
isomorphisms
then we call $(g, f)$ a bundle gerbe isomorphism.

If $J$ is a hermitian line bundle over $Y$ then we can define a
bundle gerbe $\delta(J)$ by $\delta(J) = {\pi_1^{-1}(J)}\otimes
\pi_2^{-1}(J)^*$, that is $\delta(J)_{(y_1, y_2)} = J_{y_2} \otimes J_{y_1}^*$,
where $\pi_i\,:\,Y^{[2]} \to Y$ is the map which omits the $i$th element.
The bundle gerbe product
is induced by the natural pairing
$$
J_{y_2}\otimes J_{y_1}^*\otimes J_{y_3}\otimes J_{y_2}^* \to
J_{y_3}\otimes J_{y_1}^*.$$

A bundle gerbe which is isomorphic
to a  bundle gerbe of the form $\delta(J)$ is  called {\em trivial}.
A choice of $J$ and  a bundle gerbe  isomorphism $\delta(J) \simeq L$ is called
a {\em trivialisation}.  If $J$ and $K$ are trivialisations
of $P$ then we have natural isomorphisms
$$
J_{y_1}\otimes J_{y_2}^* \simeq K_{y_1}\otimes K_{y_2}^*
$$
and hence
$$
J_{y_1}^*\otimes K_{y_1} \simeq J_{y_2}^*\otimes K_{y_2}
$$
so that the bundle $J\otimes K$ is the pull-back of a hermitian line
bundle on $M$. Moreover if $J$ is a trivialisation and
$L$ is a bundle on $M$ then $J \otimes \pi^{-1}(L)$ is
also a trivialisation.  Hence the set of all trivialisations of
a given bundle gerbe is naturally acted on by the set of all
hermitian line bundles on $M$.  This is analogous to the way in which the
set of all trivialisations of a hermitian line bundle $ L\to M$ is acted
on by the set of all maps $M \to U(1)$.

One can think of
bundle gerbes as one stage in a hierarchy of objects with
each type of object having a characteristic class in $H^p(M, \ZZ)$.
For example if $p=1$ we have maps from $M$ to $U(1)$, the characteristic
class is the pull-back of $dz$.  When $p=2$  we have hermitian line
bundles on $M$ with
characteristic class the Chern class. When $p=3$ we have bundle gerbes and
they have a characteristic class $d(L) = d(L, Y) \in H^3(M, \ZZ)$,
the Dixmier-Douady class of $(L, Y)$.  The Dixmier-Douady class is the
obstruction to the gerbe being trivial.
It is shown in \cite{Mur} that
\begin{theorem}[\cite{Mur}]
\label{th:trivial}
A bundle gerbe $(L, Y)$ has zero Dixmier-Douady class
precisely when it is trivial.
\end{theorem}

Strictly speaking the theorem in \cite{Mur} dealt with bundle gerbes defined
using line bundles or $\CC^\times$ principal bundles not hermitian
line bundles.  To see
that it generalises we need to know that if $L = \delta(J)$ for $J
\to Y$ a line
bundle then we can choose an inner product on $J$ so that $\delta(J)$ has an
isomorphic inner product to that on $L$.  Notice that if $V$ is a
one dimensional hermitian inner product then the set of vectors of
unit length is an orbit under $U(1)$. It follows that any two hermitian
inner products differ by multiplication by $e^\lambda$ for some
real number $\lambda$.  So if we choose any
hermitian inner product on the fibres of $J$ the induced inner product
on $\delta(J)$ differs from that on $L$ by a function $e^g$
where $g  \colon Y^{[2]} \to (0, \infty)$.
Because these inner products are compatible with the bundle gerbe
product we will
have that $\delta(g)(y_1, y_2, y_3) = g(y_2, y_3) - g(y_1, y_3) +
g(y_1, y_2) = 0$.
If we change the inner product on $J$ then $g$ is altered by addition
of $\delta(h)(y_1, y_2) = h(y_2) - h(y_1)$ where $h \colon Y \to
\RR$.  So we need to solve
$\delta(g) = h$ and this can be done using the exact sequence  in
Section 8 of \cite{Mur}.

Notice that the same is true of the other objects in our hierarchy,
line bundles
are trivial if and only if their chern class vanishes and maps into
$U(1)$ are
trivial (i.e. homotopic to the constant map $1$) if and only if the pull-back
of $dz$ vanishes in cohomology.

The construction
of the Dixmier-Douady class is natural in the sense that
if $Z \to N$ is another locally split map
and $\hat\phi \colon Z \to Y$ is a  fibre map covering
$\phi \colon N \to M$ then it is straightforward to
check from the definition that
\begin{equation}
\label{eq:natural0}
d(\phi^{-1}(L), Z) = \phi^*(d(L, Y)).
\end{equation}
In particular if $M = N$ and $\phi $ is the identity
then
\begin{equation}
\label{eq:natural1}
d(\phi^{-1}(L)) = d(L).
\end{equation}

          From \cite{Mur} we also have
\begin{theorem}[\cite{Mur}]
\label{th:dd}
If $L$ and $J$ are bundle gerbes over $M$ then
\begin{enumerate}
\item $d(L^*) = -d(L)$ and
\item$d(L\otimes J) = d(L)
+ d(J)$.
\end{enumerate}
\end{theorem}

\subsection{Lifting bundle gerbes}
\label{sec:lbg}
We will need one example of a bundle gerbe in a number of places.
Consider a central
extension of groups
$$
U(1) \to \hat G \to G.
$$
If $Y \to M$ is a principal $G$ bundle then it is well known that the
obstruction to lifting $Y$ to a $\hat G$ bundle is a class in $H^3(M, \ZZ)$.
It was shown in \cite{Mur} that a bundle gerbe can be constructed
from $Y$, the so-called lifting bundle gerbe, whose Dixmier-Douady
class is the obstruction to lifting $Y$ to a $\hat G$ bundle. The construction
of the lifting bundle gerbe is quite simple. As $Y$ is a principal bundle
there is a map $g \colon Y^{[2]}\to G$ defined by $y_1 g(y_1, y_2) = y_2$.
We use this
to pull back the $U(1)$ bundle $\hat G \to G$ and form the associated hermitian
line bundle $L \to Y^{[2]}$. The bundle gerbe product is induced by the group
structure of $\hat G$.

We will be interested in the lifting bundle gerbes for
$$
U(1) \to U(n) \to PU(n)
$$
and
$$
U(1) \to U(\cH)\to PU(\cH)
$$
for $\cH$ an infinite dimensional, separable, Hilbert space.

\section{Stable isomorphism of bundle gerbes}
Equation \eqref{eq:natural1} shows that there are many
bundle gerbes which have the same Dixmier-Douady class but
which  are not isomorphic.
For bundle gerbes there is a  notion called {\em stable isomorphism}
which corresponds exactly to two bundle gerbes having the
same Dixmier-Douady class.  To motivate this consider the case
of two hermitian line bundles $L \to M$ and $J \to M$ they are isomorphic
if there is a bijective map $L \to J$ preserving all structure, i.e. the
projections to $M$ and the $U(1)$ action on the fibres.  Such isomorphisms
are exactly the same thing as trivialisations of $L^*\otimes J$.  For the
case of bundle gerbes the latter is the correct notion and we have
\begin{definition}
\label{def:stableiso}
A stable isomorphism between bundle gerbes  $(L, Y)$ and $(J, Z)$
is a trivialisation of $L^*\otimes J$.
\end{definition}

We  have \cite{MurSte}

\begin{proposition}
\label{prop:stable}
A stable isomorphism exists from  $(L, Y)$ to $(J, Z)$
if and only if $d(L) = d(J)$.
\end{proposition}

If a stable isomorphism exists from $(L, Y)$ to $(J, Z)$
we say that $(L, Y)$ and $(J, Z)$ are stably isomorphic.

It follows easily that  stable
isomorphism is an equivalence relation.
It was shown in \cite{Mur} that every class in $H^3(M, \ZZ)$ is
the Dixmier-Douady class of some bundle gerbe. Hence we can deduce
from Proposition \ref{prop:stable} that
\begin{theorem}
\label{th:stableiso}
The Dixmier-Douady class defines a bijection between
stable isomorphism classes of  bundle gerbes and $H^3(M, \ZZ)$.
\end{theorem}

It is shown in \cite{MurSte} that a morphism from  $(L, Y)$ to $(J, Z)$
induces a stable isomorphism but the converse is not true.

Assume that we have a stable isomorphism $\alpha$ from  $(L, Y)$
to $(J, Z)$
and another stable isomorphism $\beta$ from $(J, Z)$ to $(K, X)$ then
it is shown
in \cite{Ste} that there is a stable isomorphism $\beta\circ \alpha$
from $(L, Y)$ to $(K, X)$ called the composition of $\alpha$ and $\beta$.
To define this we note that $d(L, Y) = d(J, Z) = d(K, X)$ so that
there exists a stable isomorphism $\gamma $ from $(L, Y)$ to $(K,
X)$. By definition
$\alpha$ is trivialisation of $L^*\otimes J$ and $\beta$ is a trivialisation
of $J^* \otimes K$. It is straightforward to show \cite{MurSte} that
$J^* \otimes J$  has
a canonical trivialisation say $\epsilon$.  Trivialisations can be
multiplied so we have
two trivialisations $\alpha\otimes \beta $ and $\gamma \otimes
\epsilon$ of $L^* \otimes J
\otimes J^* \otimes K$. It follows that there is a hermitian line bundle
$S$ over $M$ such that
$\alpha\otimes \beta = \gamma \otimes \epsilon \otimes \pi^{-1}(S)$.  We define
$\beta \circ \alpha = \gamma \otimes \pi^{-1}(S)$ or equivalently we define
$\beta \circ \alpha$ so that $\alpha\otimes \beta = (\beta \circ \alpha)
\otimes \epsilon$.  The composition of stable isomorphisms is not quite
associative, see  \cite{Ste} for details.

Notice that this construction also applies to line bundles. If $L$,
$J$ and $K$ are
line bundles over $M$ and  $\alpha $ is a section
of $L^* \otimes J$ and $\beta$ is a section of $J^* \otimes K$ then
$\beta \circ \alpha
\colon L \to K$ is a section of $L^* \otimes K$ satisfying $\alpha
\otimes \beta =
(\beta\circ \alpha) \otimes \epsilon$ where $\epsilon$ is the
canonical section of
$J^* \otimes J$.

\section{Bundle gerbe modules}

Let $(L, Y)$ be a bundle gerbe over a manifold $M$ and let $E \to Y$
be a finite rank, hermitian vector bundle. Assume that there is a
hermitian bundle
isomorphism
\begin{equation}
\label{eq:bgmod}
\phi \colon L\otimes \pi_1^{-1}E \stackrel{\sim}{\to}
\pi_2^{-1}E
\end{equation}
which is compatible with the bundle gerbe multiplication in the sense
that the two maps
$$
L_{(y_1, y_2)} \otimes (L_{(y_2, y_3)} \otimes E_{y_3} ) \to
L_{(y_1, y_2)} \otimes E_{y_2} \to E_{y_1}
$$
and
$$
(L_{(y_1, y_2)} \otimes L_{(y_2, y_3)} )\otimes E_{y_3} \to
L_{(y_1, y_3)} \otimes E_{y_3} \to E_{y_1}
$$
are the same.  In such a case we call $E$ a  bundle gerbe module and say
that the bundle gerbe acts on $E$.  Bundle gerbe modules have also
been considered for the case that $Y$ is a disjoint
union of open sets in \cite{LupUri} and in \cite{Mac} where they are called
twisted bundles.

We define two bundle gerbe modules to be isomorphic if they are
isomorphic as vector bundles and the isomorphism
preserves the action of the bundle gerbe.
Denote by $\Mod(L)$ the set of all isomorphism classes of
bundle gerbe modules for $L$.
If $(L, Y)$ acts on $E$ and also on $F$ then it acts on $E\oplus F$
in the obvious diagonal manner.  The set  $\Mod(L)$ is therefore
a semi-group.

Notice that if $E$ has rank one then it is a trivialisation
of $L$. Moreover if $E$  has  rank $r$ then $L^r$
acts on $\wedge^r(E)$ and we deduce
\begin{proposition}
\label{prop:torsion}
If $(L, Y)$ has a bundle gerbe module $Y \to E$ of rank $r$ then its
Dixmier-Douady class $d(L)$ satisfies $r d(L) = 0$.
\end{proposition}

Recall that if $E \to Y$ is a bundle then {\em descent data} \cite{Bry}
for $E$ is a collection of   hermitian isomorphisms
$\chi(y_1, y_2) \colon E_{y_2} \to E_{y_1}$ such that $\chi(y_1, y_2) \circ
\chi(y_2, y_3) = \chi(y_1, y_3)$.  The existence of descent data
is equivalent to the existence of a bundle $F \to M$ and an
isomorphism $E \to \pi^{-1}(F)$.

If $L$ is a trivial bundle gerbe then $L_{(y_1, y_2)} = K_{y_2} \otimes
K_{y_1}^*$
so if $E$ is an $(L, Y)$ module we have isomorphisms $K_{y_2} \otimes
E_{y_2} \simeq
K_{y_1} \otimes E_{y_1} $ which are descent data and hence $K \otimes E$ is the
pull back of a bundle on $M$.
     Conversely
if $F$ is a bundle on $M$ then $L$ acts on $K\otimes \pi^{-1}(F)$. Denote
by $\Bun(M)$ the semi-group of all isomorphism classes of vector
bundles on $M$. Then we have
we have
\begin{proposition}
\label{prop:trivialcase}
A trivialisation of  $(L, Y)$ defines  a semi-group
isomorphism from $\Mod(L)$ to $\Bun(M)$.
\end{proposition}

Notice that this isomorphism is not canonical but depends on the choice
of the trivialisation. If we change the trivialisation by tensoring
with the pull-back of a line bundle $J$ on $M$ then the isomorphism changes
by composition with the endomorphism of $\Bun(M)$ defined by tensoring
with the $J$.

Recall that a stable isomorphism from a bundle gerbe $(L, Y)$
to a bundle gerbe $(J, X)$ is a trivialisation of $L^*\otimes J$. This
means there is a bundle $K \to Y \times_f X$ and an isomorphism
$ L^*\otimes J \to \delta(K)$ or, in other words for
every $(y_1, y_2)$ and $(x_1, x_2)$ we have an isomorphism
$$
     L_{(y_1, y_2)}^*\otimes J_{(x_1, x_2)}
\to K_{(y_2, x_2)} \otimes K_{(y_1, x_1)}^*.
$$

Let $E \to Y$ be an $L$ module and define $\hat F_{(y, x)} = K_{(y,x)}^* \otimes
E_y$ a bundle on $Y \times_f X$.   We have isomorphisms
\begin{align*}
\hat F_{(y_2, x)} &= K_{(y_2,x)}^* \otimes E_{y_2}\\
                      &= K_{(y_1, x)}^* \otimes L_{(y_1, y_2)} \otimes 
E_{y_2} \\
                      &= K_{(y_1, x)}^* \otimes E_{y_1} \\
                      &= \hat F_{(y_1, x)}.
\end{align*}
These define   a descent map
for $\hat F$ for the map $Y \times_\pi X \to X$
and hence define a bundle $F$ on $X$.  Note that
as the inner products are everywhere preserved $F$ is also
a hermitian bundle.

We also have
\begin{align*}
J_{(x_1, x_2)} \otimes F_{x_2} &=J_{(x_1, x_2)} \otimes K_{(y,
x_2)}^*\otimes E_y \\
                         &= K_{(y, x_1)}^* \otimes E_y \\
                         & = F_{x_1}
\end{align*}
and this makes $F$ a $(J, X)$ module.

So the choice of stable isomorphism has defined a map
$$
\Mod(L) \to \Mod(J).
$$
In a similar fashion we can define a map
$$
\Mod(J) \to \Mod(L)
$$
which is an inverse.  Hence we have

\begin{proposition}
\label{prop:module}
A stable isomorphism from $(L, Y)$ to
$(J, Y)$ induces an isomorphism of semi-groups
between $\Mod(L)$ and $\Mod(J)$.
\end{proposition}

Note that, as in the trivial case, this isomorphism is not
canonical but depends on the choice of stable isomorphism. Changing the
stable isomorphism by tensoring with the pull-back of a line bundle
$J$ over $M$
changes the isomorphism in Prop.  \ref{prop:module}
by composition with the endomorphism
of $\Mod(J)$ induced by tensoring with the pull-back of $J$.

There is a close relationship between bundle
gerbe modules and bundles of projectives spaces. Recall that a
bundle of projective spaces $\cP \to M$ is a fibration whose fibres
are isomorphic
to $P(V)$ for $V$ a Hilbert space, either finite or infinite dimensional,
and whose structure group is $PU(V)$. This  means that there is a
$PU(V)$ bundle $X \to M$ and $\cP = X \times_{PU(V)} P(V)$.
Associated to $X$ is a
lifting bundle gerbe $J \to X^{[2]}$ and a Dixmier-Douady class.
This Dixmier-Douady class is the obstruction to $\cP$ being the
projectivisation of a vector bundle.  The lifting bundle gerbe acts naturally
on the bundle gerbe module $E = X \times H$ because each $J_{(x_1,
x_2)} \subset U(V)$ by
construction \ref{sec:lbg}.

Let $(L, Y)$ be a
bundle gerbe and $E \to Y$ a bundle gerbe module. Then the projectivisation
of $E$ descends to a projective bundle $\cP_E \to M$ because of the
bundle gerbe action. It is straightforward to check that the class of this
projective bundle is $d(L)$.  Conversely if $\cP \to M$ is a projective
bundle with class $d(L)$ the associated lifting bundle gerbe has class
$d(L)$ and hence is stably isomorphic to $(L, Y)$. So the module on which the
lifting bundle gerbe acts defines a module on which $(L, Y)$ acts.  From the
discussion before \ref{prop:module} one can see that if  two modules
are related
by a stable isomorphism they give rise to the same projective bundle
on $M$.   We also  have that $E \to Y$ and $F \to Y$ give rise to
   isomorphic projective bundles on $M$ if and only if there is a line
bundle $K \to M$ with $E = \pi^{-1}(K)\otimes F$.
Denote by $\Lin(M)$ the group
of all isomorphism classes of line bundles on $M$. Then this acts on
$\Mod(L)$ by $E \mapsto \pi^{-1}(K) \otimes E$ for any line
bundle $K \in \Lin(M)$.  If $[H] \in H^3(M, \ZZ)$ denote
by $\Pro(M, [H])$ the set of all isomorphism classes of
projective bundles with class $[H]$.
Then we have

\begin{proposition}
\label{prop:proj}
If $(L, Y)$ is a bundle gerbe then the  map
which associates to any element of $\Mod(L)$ a projective
bundle on $M$ whose Dixmier-Douady class is equal to $d(L)$ induces
a bijection
$$
\frac{\Mod(L)}{\Lin(M)} \to \Pro(M, d(L)).
$$
\end{proposition}

\section{$K$-theory for torsion bundle gerbes}

Given a bundle gerbe $(L, Y)$ with torsion Dixmier-Douady class
we denote by
$K(L)$ the Grothendieck group of the
semi-group $\Mod(L)$ and call this the
$K$ group of the bundle gerbe. We  immediately have
from Prop. \ref{prop:module}:

\begin{proposition}
             A choice of stable isomorphism from $L$ to $J$ defines a
             canonical isomorphism $K(L) \simeq K(J)$.
             \end{proposition}

Notice that the group $K(L)$ depends only on the class $d(L) \in H^3(M, \ZZ)$
and for any class $[H]$ in $H^3(M, \ZZ)$ we can define a bundle gerbe $L$ with
$d(L)= [H] $ and hence a group $K(L)$. When we want to emphasise the
dependence on $[H]$ we denote
this by $K_{bg}(M, [H])$.

It is easy to deduce from the theory of bundle gerbes various
      properties of this $K$-theory:

\begin{proposition}
\label{prop:bgkprops}
Bundle gerbe $K$ theory satisfies the following properties:\\

(1) If $(L, Y)$ is a trivial bundle gerbe then $K_{bg}(L) = K(M)$.

(2) $K_{bg}(L)$ is a module over $K(M)$.

(3)  If $[H]$ and $[H']$ are classes in $H^3(M, \ZZ)$ there is a homomorphism
$$
K_{bg}(M, [H]) \otimes K_{bg}(M, [H']) \to K_{bg}(M, [H] + [H']).
$$

(4) If $[H]$ is a class in $H^3(M, \ZZ)$ and $f \colon N \to M$ is a
map  there is a homomorphism
$$
K_{bg}(M, [H]) \to K_{bg}(N, f^*([H])).
$$

\end{proposition}
\begin{proof}
(1)   This follows from applying
Prop.~\ref{prop:trivialcase} which shows that
$\Mod(L)$ is isomorphic to the
semi-group of all vector bundles on $M$

(2) If we pull a bundle back from $M$ to $Y$ and tensor it with a bundle
gerbe module the result is still a bundle gerbe module.

(3) If $E \to Y$ is a bundle gerbe module for $(L, Y)$ and $F \to X$
is a bundle gerbe module for $(J, X)$ it is straightforward to see that
$E \otimes F$ defines a bundle over the fibre product of $Y$ and $X$
which is a bundle gerbe module for $L \otimes J$.

(4) This follows easily by pull-back.

\end{proof}

There is another  construction that associates to any class
$[H]$ in  $H^3(M, \ZZ)$ a group $K(M, [H])$ or the  {\em twisted}
$K$ group. Twisted
$K$-theory shares the same properties as those in Prop. \ref{prop:bgkprops}.
In the next section  we discuss twisted cohomology and show that,
in the torsion case, bundle gerbe cohomology and twisted cohomology
coincide.

\section{Twisted $K$-theory and bundle gerbe modules}

\subsection{Twisted $K$-theory}

We recall the definition  of twisted cohomology  \cite{Ros}.
In this discussion the class $[H] \in H^3(M, \ZZ)$ is
not restricted to be torsion.

Given a class $[H] \in H^3(M, \ZZ)$ choose a $PU(\cH)$ bundle $Y$ whose
class is $[H]$. We can form an associated bundle
$$
Y(\Fred) = Y \times_{PU(\cH)} \Fred
$$
where $\Fred$ is the space of Fredholm operators on $\cH$ acted
on by conjugation.  Let $[M, Y(\Fred)]$ denote the
space of all homotopy classes of sections of $Y(\Fred)$ then we have
\cite{Ros}
\begin{definition}[\cite{Ros}]
\label{def:twisted}
If $[H] \in H^3(M, \ZZ)$ the twisted $K$ theory
of $M$ is defined by
$$
K(M, [H]) = [M, Y(\Fred)].
$$
\end{definition}

It is a standard result that sections of $Y(\Fred)$ are equivalent
to $PU(\cH)$ equivariant maps from $Y \to \Fred$ so we have
\begin{equation*}
K(M, [H]) =[M, Y(\Fred)]_{PU(\cH)}
\end{equation*}
where $[M, Y(\Fred)]_{PU(\cH)}$ is the space
of all homotopy classes of equivariant maps with the homotopies
being by equivariant maps.

\subsection{Bundle gerbe  $K$-theory and twisted $K$-theory in the
torsion case}

In the case when the Dixmier-Douady class $[H]$ is torsion, we will
prove that bundle gerbe $K$ theory and twisted $K$-theory
are the same and indicate their relationship with equivariant
$K$-theory.

The Serre-Grothendieck theorem cf. \cite{DK} says that, given a
torsion class, there is a $PU(n)$ bundle $X\to M$, with Dixmier-Douady
invariant equal to $[H]$. We can define an action  $U(n)$ on $\CC^n
\otimes \cH = \cH^n$
letting $g$ act as $g \otimes 1$. This gives a
representation $\rho_n \colon U(n) \to U(\cH^n)$
and induces  a $PU(\cH^n)$ bundle with Dixmier-Douady
   class $[H]$. As $\cH^N \simeq \cH$ and all $PU(\cH)$
bundles are determined by their Dixmier-Douady class we can
assume that this bundle is $Y$ and contains $X$ as a $U(n)$
reduction.  Then we have
$$
(Y \times \Fred)/PU(\mathcal H)  \cong (X \times \Fred)/PU(n),
$$
so that
$$
K(M,  [H]) = [Y, \Fred]^{PU(\mathcal H)} \cong [X, \Fred]^{PU(n)}.
$$

The lifting bundle gerbe for $Y \to M$ pulls-back to become the lifting
bundle gerbe $L$ for $X \to M$.  We will now prove that $K_{bg}(M, L)
= K(M, [H])$. Notice
that this will prove the result also for {\em any} bundle gerbe with
torsion Dixmier-Douady class as we already know that bundle gerbe K-theory
depends only on the Dixmier-Douady class.

In the case where there is no twist Atiyah showed that $K(M) = [M,
\Fred]$ and we
will follow his proof indicating just what needs to be modified to
cover this equivariant case.

First we have have the following
\begin{lemma} If $W$ is  a finite dimensional subspace of $\CC^n
\otimes \cH$ there
is a finite co-dimensional subspace $V$ of $\cH$ such that $\CC^n \otimes V
\cap W = 0$.
\end{lemma}
\begin{proof} Let $U$ be  the image of $V$ under the map $\CC^n
\otimes \CC^n \otimes
\cH \to \cH$ were we contract the two copies of $\CC^n$ with the inner
product. Then $V \subset \CC^n \otimes U$. So take $W = U^{\perp}$.
\end{proof}

Using the compactness of $X$ and the methods in Atiyah we can show that if
$f \colon X \to \Fred(\CC^n \otimes \cH)$ then there is a
   subspace $V \subset \cH$, of finite co-dimension, such that
$\ker(f(x)) \cap \CC^n \otimes V = 0$. Then $\cH/V$ and $\cH/f(V)$
will be vector bundles
on $X$ and moreover they will  by acted on by $U(n)$ in such a way as to make
them bundle gerbe modules.  So we define
$$
\ind \colon [X, \Fred(\CC^n \otimes \cH ) ] _{U(n)}\to K_{bg}(M, L)
$$
by $\ind(f) = \cH/V -\cH/f(V)$. Again the methods of \cite{Ati} will show that
this index map is well-defined and a homomorphism.

As in \cite{Ati} we can identify the kernel of $\ind$ as
$ [X, U(\CC^n \otimes \cH ) ] _{U(n)}$ and use the result of Segal \cite{Seg2}
which shows that $U(\CC^n \otimes \cH )$ is contractible so $\ind$ is
injective.

Finally we consider surjectivity. First we need from \cite{Seg1} the following
\begin{proposition}
\label{prop:sub}
If $E \to X$ is a bundle gerbe module for $L$ then there is a representation
$\mu\colon U(n) \to U(N)$ such that $E$ is a sub-bundle gerbe module of
$\CC^N \otimes X$.
\end{proposition}

If $E \to X$ is a bundle gerbe module then Proposition \ref{prop:sub}
enables us to find a $U(n)$
equivariant map $\tilde f \colon X \to \Fred(\CC^N \otimes X)$ whose
index is $E$.
The action of $U(n)$ used here on $\CC^N \otimes X$ is that induced from the
representation $\mu$.
To prove surjectivity of the index map it suffice to find a
map $f \colon X \to \Fred(\CC^n \otimes X)$
whose index is $E$. Then if $E-F$ is a class in $K_{bg}(M, L)$ we can
apply a similar technique
to obtain a map whose index is $-F$ and combine these to get a map
whose index is $E-F$ and we are done.

To construct $f$ we proceed as follows.  We have a representation
$\rho_n \colon U(n) \to \CC^n \otimes \cH$
and a representation  $\rho_N \colon U(n) \to \CC^N \otimes \cH$.
These can be used to induce a
$PU(\CC^n \otimes \cH)$ bundle and a $PU(\CC^N \otimes \cH)$ bundle, both
with Dixmier-Douady class $[H]$.
So they must be isomorphic.  We need the precise form of this isomorphism.
Choose an isomorphism $\phi \colon \CC^n \otimes \cH \to \CC^N
\otimes \cH$. This induces
an isomorphism $U(\CC^n \otimes \cH )\to U(\CC^N \otimes \cH)$ given
by $u \mapsto
\phi u \phi^{-1}$ which we will denote by $\phi[u]$ for convenience.
There is a similar
identification $\Fred(\CC^n \otimes \cH) \to \Fred(\CC^N \otimes \cH)$. The two
$PU$ bundles are given by $X \times_{\rho_n} U(\CC^n \otimes \cH)$
and $X \times_{\rho_N} U(\CC^N \otimes \cH)$
and consist of cosets  $[x, u] = [xg, \rho^{-1}_n(g)u]$ and $[x, u] =
[xg, \rho^{-1}_N(g)u]$
respectively. The action of
$U(\CC^n \otimes \cH)$ is $[x, u]v = [x, uv]$ and similarly for
$U(\CC^N \otimes \cH)$. Because these
are isomorphic bundles there must be a bundle map
$$
\phi \colon X \times_{\rho_n} U(\CC^n \otimes \cH) \to
\times_{\rho_N} U(\CC^N \otimes \cH)
$$
satisfying $\phi([x,u]v) = \phi([x, u]) \phi[v]$ and hence
$\phi([x,u]) = \phi([x, 1]) \phi[u]$.
Define $\alpha \colon X \to U(\CC^N \otimes \cH)$ by requiring that
$\phi([x, 1]) =
[x, \alpha(x)]$. Then if $g \in U(n)$ we have
\begin{align*}
[xg, \alpha(xg)] &= \phi([xg, 1])\\
& = \phi([x, \rho_n(g)]) \\
&= \phi([x, 1])\phi[\rho_n(g)] \\
&= [x, \alpha(x)]\phi[\rho_n(g)] \\
&= [x, \alpha(x)\phi[\rho_n(g)]] \\
&= [xg, \rho_N(g)^{-1} \alpha(x) \phi[\rho_n(g)]]
\end{align*}
so that
\begin{equation}
\label{eq:relation}
\alpha(xg) = \rho_N(g)^{-1} \alpha(x) \phi[(\rho_n(g)].
\end{equation}

We can now define $ f \colon X \to \Fred(\CC^n \otimes X)$ by
$$
f(x)  =     \alpha(x)\phi^{-1}[ \tilde f(x) ]
$$
and it is straightforward to see that this is $U(n)$ equivariant
by applying the equation \eqref{eq:relation}.  It is clear that
$\alpha(x)$ and $\phi$ can be used to
establish an isomorphism between $\ker(f)$
and $\ker(\tilde f)$ and hence between $\ker(f) $ and $E$.

This proves
\begin{proposition} If $L$ is a bundle gerbe over $M$ with
Dixmier-Douady class $[H]$
which is torsion then
$K_{bg}(M, L) = K(M, [H])$.
\end{proposition}

The lifting bundle gerbe for $Y \to M$ pulls-back to become the lifting
bundle gerbe for $X \to M$.  A bundle gerbe module for
this is a bundle $E \to X$ with a $U(n)$ action covering the
action of $PU(n)$ on $X$.  This $U(n)$ action has to have the
property that the center $U(1) \subset U(n)$ acts on the fibres of
$E$ by scalar multiplication.  Considered from this perspective we see that
we are in  the context
of equivariant $K$ theory \cite{Seg1}. Notice that by projecting to
$PU(n)$ we can make
$U(n)$ act on
$X$. Of course this action is
not   free, the center $U(1)$ is the isotropy subgroup
at every point.   The equivariant $K$ theory
$K_{U(n)}(Q)$ is the $K$ theory formed from vector bundles
on $Q$ which have an $U(n)$ action covering the action on $X$.  We
need a subset
of such bundles with a particular action.  To understand this note that
because the center $U(1) \subset U(n)$ is the isotropy subgroup for the $U(n)$
action on $X$ it must act on the fibres of $E$ and hence define a
representation
of $U(1)$ on $\CC^r$ if the bundle $E$ has rank $r$.  This defines an
element of
$R(U(1))$, the representation ring  of $U(1)$, and to be a bundle gerbe module
this representation must be scalar multiplication on $\CC^r$.  In terms
of equivariant $K$ theory we can consider the map which is
restriction to a fibre of $X \to M$ and then we have
$$
K_{U(n)}(X) \to K_{U(n)}(U(n)/U(1)) = R(U(1)).
$$
$K_{bg}(M, L)$ is the pre-image under this map of the representation
of $U(1)$ on $\CC^n$ by scalar multiplication.

%:Characteristic classes of bundle gerbe modules

\subsection{Characteristic classes of bundle gerbe modules}
In this section we discuss the Chern character of a twisted
bundle gerbe module.
Suppose $(L,Y)$ is a bundle gerbe on $M$ and
that $E\to Y$ is a bundle gerbe module. Recall \eqref{eq:bgmod}
that this means that there is an isomorphism:
$$
\phi \colon L\otimes \pi_1^{-1}E \ \stackrel{\sim}{\to}\
\pi_2^{-1}E
$$
which is compatible with the bundle gerbe product
on $L$.  Recall from \cite{Mur} that a \emph{bundle
gerbe connection} on $L$ is a connection $\nabla_L$
on $L$ which is compatible in the obvious sense with
the bundle gerbe product on $L$.  Furthermore, one
can show (see \cite{Mur}), that the curvature
$F_L$ of $\nabla_L$ satisfies $F_L = \delta(f) = \pi_1^*f
- \pi_2^*f$ for some $2$-form $f$ on $Y$.  $f$ is unique
up to $2$-forms pulled back from $M$.  We call a choice
of such an $f$ a \emph{curving} for the connection
$\nabla_L$.  In \cite{Mur} it is shown that there is
a closed, integral $3$-form $\omega$ on $M$ such
that $df = \pi^* \omega$.  $\omega$ is called the
$3$-curvature of the connection $\nabla_L$ and
curving $f$.  It is the image in real cohomology
of the Dixmier-Douady class of $L$.  In our case,
since $L$ has torsion Dixmier-Douady class, one
can choose a bundle gerbe connection $\nabla_L$
for $L$ and a curving $f$ for $\nabla_L$ such that
$df =0$.
We want a connection $D$ on
$E$ so that $\phi$ is a connection preserving
isomorphism of vector bundles, where $L\otimes
\pi_1^{-1}E$ is given the tensor product connection
$\nabla_L \otimes I + \pi_1^{-1}D$.

Take an open cover $\{U_i\}_{i\in I}$ of $M$ such
that there exist local sections over $U_i$ of
$\pi\colon Y\to M$ and such that there exists a
partition of unity $\{\rho_i\}_{i\in I}$ of $M$
subordinate to $U_i$.  Then $L$ is trivialised over
$U_i$ --- say $L = \delta(K_i)$ over $U_i$.  The connection
$\nabla_L$ on $L$ induces a connection $\nabla_i$ on
$K_i$.  The bundle $E\otimes K_i$ on $Y_i = Y|_{U_i}$ descends
to a bundle $F_i$ on $U_i$.  Choose any connection
$\nabla_E$ on $E$, and a connection $\nabla_{F_i}$ on
$F_i$.  Then the pullback connection $\pi^{-1}\nabla_{F_i}$
on $\pi^{-1}F_i$ differs from the connection
$\nabla_E + \nabla_i\otimes I$ on $E\otimes K_i$
by an $\End(E\otimes L_i) = \End(E)$ valued $1$-form
$B_i$ on $Y_i$.  Give $E|_{Y_i}$ the connection
$D_i = \nabla_E - B_i$.  Then, over $Y^{[2]}_i$, $\phi$
induces an isomorphism of vector bundles with
connection
$$
L|_{Y_i^{[2]}}\otimes \pi_1^{-1}E|_{Y_i^{[2]}}
\stackrel{\sim}{\to} \pi_2^{-1}E|_{Y_i^{[2]}}.
$$
Using the partition of unity $\{\rho_i\}_{i\in I}$
pulled back to $\{Y_i\}_{i\in I}$ we can patch together
the local connections $D_i$ on $E|_{Y_i}$ to get a
connection $D$ on $E$ which is compatible with $\nabla_L$ under
$\phi$ in the above sense.
Calculating curvatures we get the following equality
of $\End(\pi_1^{-1}E\otimes L) = \End(\pi_2^{-1}E)$
valued $2$-forms on $Y^{[2]}$:
\begin{equation} 
\label{eq:module connection transfm law}
F_{\pi_1^{-1}D} + F_L I = \phi^{-1}\circ F_{\pi_2^{-1}D}
\circ \phi.
\end{equation} 
Writing $F_L = \pi_1^* f - \pi_2^* f$ we get
$$
\pi_1^*(F_{D} + f I) = \phi^{-1}\circ \pi_2^*(F_{D} + f I)\circ \phi.
$$
   If $P$ is an invariant polynomial in
Lie algebra valued variables then this
equation shows that $$
\pi_1^*(P(F_D + fI,\ldots ,F_D + fI)) = \pi_2^*(P(F_D + fI,\ldots ,F_D + fI))
$$
and hence the Chern-Weil $2k$-forms
on $Y$ descend to $M$.  Moreover
\begin{eqnarray*}
&   & dP(F_D+fI,\ldots ,F_D+fI) \\
& = & \sum P(F_D+fI,\ldots ,dF_D + dfI,\ldots ,F_D+fI) \\
& = & \sum P(F_D+fI,\ldots,[F_D,A],\ldots ,F_D+fI).
\end{eqnarray*}
Using the standard trick of writing $g_t = \exp (tA)$
and using the invariance of $P$ we get
\begin{eqnarray*}
0 & = & \frac{d}{dt}|_{t=0} P(g_t^{-1}F_Dg_t + fI,\ldots,
g_t^{-1}F_Dg_t + fI)                                       \\
& = & \sum P(F_D+fI,\ldots ,[F_D,A],\ldots ,F_D+fI).
\end{eqnarray*}
So the $2k$-forms $P(F_D+fI,\ldots ,F_D+fI)$ on
$M$ are all closed.

The usual definition of the chern character can be applied to define
$\Ch(E) \in H^*(M, \QQ)$
for any bundle gerbe module. This satisfies $\Ch(E+F) = \Ch(E) +
\Ch(F)$ and hence
defines a chern character:
$$
\Ch \colon K_{bg}(M) \to H^*(M, \QQ).
$$
% ---------------------------------------------------------------------

%:Twisted K-theory in the non-torsion case
\section{Twisted $K$-theory in the  non-torsion case.}
\label{sec:non-torsion case} 

We have seen that approaching twisted $K$-theory via finite rank
bundle gerbes is not possible if the class $[H]$
is not torsion as there are then no finite rank bundle gerbe modules.
A possible
generalisation would be to allow bundle gerbe modules which
are infinite Hilbert bundles. In that case the induced
projective bundle on $M$ is a $PU(\cH)$ bundle for $\cH$
an infinite dimensional Hilbert bundle and it is well
known that there is only one such bundle for a given
Dixmier-Douady class and hence Proposition \ref{prop:proj}
implies that

\begin{proposition}
Every bundle gerbe admits exactly one bundle gerbe module which is a
bundle of infinite dimensional Hilbert spaces  with structure group $U(\cH)$.
\end{proposition}

In particular if $E$ and $F$ are Hilbert bundle gerbe modules
then $E = F$ so that the class $E=F$ in the induced
$K$ group is zero. So the $K$ group is zero.

In the remainder of this section
we discuss another approach to twisted cohomology
where the structure group of the bundle gerbe module
is the group $\UK$, the subgroup of $U(\cH)$ 
of unitaries which differ from the identity by a compact 
operator (here $\mathcal{K}$ denotes the compact operators 
on $\cH$).  
To see how this arises notice that in
Rosenberg's definition \ref{def:twisted} we can replace  $\Fred$ by a
homotopy equivalent space.
For our purposes we choose $BU{_\mathcal K}\times \ZZ$.  
This can be done in a $PU(\cH)$
equivariant fashion as follows. For $BU{_\mathcal K}$
we could choose the connected component of the
identity of the invertibles in the Calkin algebra $B(\cH)/\mathcal K$
which is homotopy equivalent in a $PU(\cH)$ equivariant way
to the Fredholms of index zero under the quotient map
$\pi: B(\cH)\to B(\cH)/\mathcal K$. Note however
that the identity component of the invertibles in the Calkin
algebra is just $GL(\cH)/GL_\mathcal K$ where $GL(\cH)$
denotes the invertible operators on $\cH$ and $GL_\mathcal K$
are the invertibles differing from the identity by a compact.
Thus we can take $BU_{\mathcal K}$ to be $GL(\cH)/GL_{\mathcal K}$
and this choice is $PU(\cH)$ equivariant. We could equally well take
$B\UK = U(\cH)/\UK$.

As $U(\cH)$ acts on $U_{\mathcal K}$ by conjugation there is
a semi-direct product
$$
U_{\mathcal K} \to U_{\mathcal K} \rtimes PU(\cH) \to PU(\cH).\
$$
Note that this means that any $ U_{\mathcal K} \rtimes PU(\cH)$
bundle over $M$ induces a $PU(\cH)$ bundle and hence a class
in $H^3(M, \ZZ)$.   If $R \to Y$ is a $U_{\mathcal K}$ bundle
we call it $PU(\cH)$ covariant if there is an action
of $PU(\cH)$ on the right of $R$ covering the action on $Y$ such that
$(rg)[u] = r[u] u^{-1}gu$ for any $r \in R$,
$[u] \in PU(\cH)$ and $g \in U_{\mathcal K}$. Here $[u]$ is the
projective class of some $u \in U(\cH)$.

Because $B\UK$ is homotopy equivalent to only the connected component
of index $0$ of $\Fred$ it is convenient to work with reduced twisted
$K$ theory, $\redK(M, [H])$, defined by
$$
\redK(M, [H]) = [M, Y(B\UK)].
$$

We have
\begin{proposition}
Given a $PU(\cH)$ bundle $Y \to M$ with class $[H] \in H^3(M, \ZZ)$ the
following are equivalent
spaces
\begin{enumerate}
\item $\redK(M, [H])$
\item space of homotopy classes of sections of $Y \times_{PU(\cH)}
BU{_\mathcal K}$
\item space of homotopy classes of  $PU(\cH)$ equivariant maps from $Y$ to
$BU{_\mathcal K}$
\item space of isomorphism classes of $PU(\cH)$ covariant $U{_\mathcal
K}$ bundles on $Y$, and
\item space of isomorphism classes of $U{_\mathcal K} \rtimes PU(\cH)$
bundles on $M$ whose projection to a $PU(\cH)$ bundle has class $[H]$.
\end{enumerate}
\end{proposition}

\begin{proof}
(1) $\iff$ (2) This is just the reduced version of Rosenberg's
definition of twisted $K$-theory \ref{def:twisted}.

(2) $\iff$ (3) This is a standard construction.

(3) $ \implies$ (4) Notice that the $\UK$ bundle $E\UK \to B\UK$ is $PU(\cH)$
covariant. It follows that if we pull it back to $Y$ by a $PU(\cH)$
equivariant map $Y \to B\UK$ that we must get a
$PU(\cH)$ covariant bundle on $Y$.

(4) $ \iff$ (5) Let $R \to Y$ be a $PU(\cH) $ covariant $\UK$ bundle.
By composing the projections $R \to Y \to M$ we think of $R$ as a
bundle on $M$. Both groups $\UK$ and $PU(\cH)$ act on $R$ and the
combined action is an action of the semi-direct product and realises
$R$ as a bundle over $M$ for this semi-direct product.  Conversely
consider a bundle $R \to M$ for the semi-direct product for which the
induced $PU(\cH)$ bundle is isomorphic (as a $PU(\cH)$ bundle) to $Y$.
Identify this bundle with $Y$ and hence $R$ is a bundle over $Y$ and,
in fact, a $PU(\cH)$ covariant $\UK$ bundle.

(5) $ \implies$ (3) A $\UK \rtimes PU(\cH)$ bundle over $M$ is
determined by a classifying
map $\phi \colon M \to B(\UK \rtimes PU(\cH))$.  A little thought shows
that we can realise
this latter space as $EPU(\cH) \times_{PU(\cH)} B\UK$ which fibres over
$BPU(\cH)$. The composition
$\tilde\phi \colon M \to BPU(\cH)$ of $\phi$ with the projection to
$BPU(\cH)$ is the
classifying map of the induced $PU(\cH)$ bundle which is $Y$. This
means that we can find
a $PU(\cH)$ equivariant map $\hat\phi \colon Y \to EPU(\cH)$
covering $\tilde\phi$.
Using this we define $\rho \colon Y \to B\UK$ by $\phi(\pi(y)) = [
\hat\phi(y), \rho(y)]_{PU(\cH)}$.
This is well-defined. Moreover if $g \in PU(\cH)$
then $\pi(yg) = \pi(y)$ so that
$ [ \hat\phi(y), \rho(y)]_{PU(\cH)} = [\hat\phi(yg), \rho(yg)]_{PU(\cH)}  =
        [\hat\phi(y)g, \rho(yg)]_{PU(\cH)} = [\hat\phi(yg),
g\rho(yg)]_{PU(\cH)}$ and hence
$\rho(yg) = g^{-1} \rho(y)$ proving equivariance.

\end{proof}

\begin{note} Notice that if we worked with $\Fred$ instead
of $B\UK$ then it has connected components $\Fred_n$ consisting
of operators of index $n$.  We can then consider sections
of $Y \times_{PU(\cH)} \Fred_n$ for every $n$, not just zero.
Such a section will pull back a $K$ class and if we take
the determinant of this $K$ class we will obtain a line bundle
on $Y$ on which the gerbe $L^n$ acts.  Hence we will have
$n [H] = n d(L) = 0$ as in Prop. \ref{prop:torsion} and so
we deduce the result noted in \cite{Ati2} that if $[H] $ is
not torsion then there are no sections of $Y \times_{PU(\cH)} \Fred_n$
except when $n = 0$ so $\redK(M, [H]) = K(M, [H])$.
\end{note}

\subsection{$\UK$ bundle gerbe modules}

Given a $PU(\cH)$ covariant $U{_\mathcal K}$ bundle $R$ over $Y$ we can
define the associated bundle
\begin{equation}
\label{eq:associated}
E = R \times_{U{_\mathcal K}} \cH \to Y.
\end{equation}
We claim that this is a bundle gerbe module for the
lifting bundle gerbe $P$.
Let $[r, v] \in E_{y_1}$ be a $U{_\mathcal K}$ equivalence class
where $r \in R_{y_1}$, the
fibre of $R$ over $y_1 \in Y$  and $v \in \cH$.
Let $u \in L_{y_1y_2}$ be an element of the lifting bundle
gerbe. Then, by definition, $u \in U(\cH)$ and $y_1[u] = y_2$.  We
define the action of $u$ by $[r, v]u = [r[u], u^{-1}v]$.  It is
straightforward to check that this is well defined.  Hence we have
associated to any $PU(\cH)$ covariant $U{_\mathcal K}$ bundle $R$ on $Y$
a module for the lifting bundle gerbe.

The inverse construction is also possible if the bundle gerbe
module is a $\UK$ bundle gerbe module which we now define.
Let $E\to Y$ be a  Hilbert bundle with structure
group $\UK$.  We recall what it means for a Hilbert bundle
to have structure group $\UK$.  To any Hilbert bundle  there is  associated a
$U(\cH)$ bundle
$U(E)$ whose fibre, $U(E)_y$, at $y$ is all unitary isomorphisms
$f \colon \cH \to E_y$. If
$u \in U(\cH)$ it acts on $U(\cH)_y$ by $fu = f \circ u$ and hence $U(E)$ is
a principal $U(\cH)$ bundle.  For $E$ to have structure
group $\UK$ means that we have a reduction of $U(E)$ to a $\UK$ bundle
$R \subset U(E)$.
Each  $R_y \subset U(E)_y$ is an orbit under $\UK$, that
is $R$ is a principal $\UK$ bundle.

For $E$ to be a $\UK$ bundle gerbe module we need to define
an action of the bundle gerbe on it.  By comparing with the action on the
bundle $E$ defined in \eqref{eq:associated} we see that we need to make the
following  definition.  If
$u \in U(\cH)$ such that $y_{1}[u] = y_{2}$  then $u \in L_{(y_1, y_2)}$
where $L \to Y^{[2]}$ is the lifting bundle gerbe so if $f \in R_{y_1}$ then
$ufu^{-1} \in U(E)_{y_2}$.  We require that $ufu^{-1} \in R_{y_2}$. So
a lifting bundle gerbe module which is a $\UK$ Hilbert bundle and
satisfies this condition we call a $\UK$ bundle gerbe module.  By
construction
we have that the associated $R$ is a $\UK$ bundle over $Y$ on which
$PU(\cH)$ acts.  Let us denote by $\Mod_{\UK}(M, [H])$
the semi-group of all $\UK$ bundle gerbe modules for the
lifting bundle gerbe of the $PU(\cH)$ bundle with three class $[H]$. As
any two $PU(\cH)$ bundles with the same three class are isomorphic we see
that $\Mod_{\UK}(M, [H])$ depends only on $[H]$.

We have now proved
\begin{proposition}
\label{prop:U_K bg modules equals K(M;H)} 
If $(L, Y)$ is the lifting bundle gerbe for a $PU(\cH)$
bundle with Dixmier-Douady class $[H]$
$$
\redK(M, [H]) =  \Mod_{\UK}(M, [H]) .
$$
\end{proposition}

If $L_1$ and $L_2$ are two $PU(\cH)$ covariant $U{_\mathcal K}$ bundles
on $Y$ note that
$L_1 \times L_2$ is a $U{_\mathcal K} \times U{_\mathcal K}$ bundle.
Choose an
isomorphism $\cH \times \cH \to \cH$ which induces an isomorphism
$U{_\mathcal K} \times U{_\mathcal K} \to U{_\mathcal K}$ and hence defines a
new $PU(\cH)$ covariant bundle $L_1 \otimes L_2$.  It is straightfoward
to check that
$$
(L_1 \otimes L_2 )(\cH) \simeq L_1(\cH) \times L_2(\cH).
$$

This makes $\Mod_{\UK}(M, [H])$ a semi-group and the map
$\redK(M, [H]) =  \Mod_{\UK}(M, [H])$ is a semi-group isomorphism.
Note that with our definition $B\UK$ is a group. Moreover
the space of all equivariant maps $Y \to B\UK$ is a group as well.
To see this notice that if
   $f$ and $g$ are equivariant maps and we multiply
pointwise then for  $y \in Y$ and $[u] \in PU(\cH)$ we have
\begin{align*}
(fg)(y[u]) &= f(y[u]) g(y[u]) \\
            &= (u^{-1} f(y) u) (u^{-1} g(y) u) \\
            &= u^{-1}(fg)(y) u
\end{align*}
and if $f^{-1}$ is the pointwise inverse then  $f^{-1}(y[u]) =
\left( u^{-1} f(y) u \right)^{-1} = u^{-1} f^{-1}(y) u$.
This induces a group structure on $\Mod_{\UK}(M, [H])$. We have already
noted that it is a semi-group but this implies more, for every $\UK$ bundle
gerbe module $E$ there is a $\UK$ bundle gerbe module  
$E^{-1}$ such that $E \oplus E^{-1}$
is the trivial $\UK$ bundle gerbe. Hence we have

\begin{proposition} If $(L, Y)$ is the lifting bundle gerbe for a $PU(\cH)$
bundle with Dixmier-Douady class $[H]$ then
$$
    K_{\UK}(M, [H]) = \Mod_{\UK}(M, [H])  = \redK(M, [H])
$$
\end{proposition}

\begin{note}
\label{note:replace U_K with U_1} 
(1) The group $\UK$ used here could be replaced by any
other group to which it is homotopy equivalent by a homotopy
equivalence preserving the $PU(\cH)$ action.  In particular we 
could consider $U_1$, the subgroup of $U(\cH)$ consisting 
of unitary operators which differ from the identity by a 
trace class operator.   
In Section 9 we show that the computation in Section 6 of bundle gerbe
characteristic classes generalizes, with some modifications, to
$U_1$ bundle gerbe modules.  

(2) In \cite{Har} it is argued that $\UK$ is the appropriate gauge
group for non-commutative gauge theory.
\end{note}

% ------------------------------------------------------------------------
\subsection{Local description of $\UK$ bundle gerbe modules}
Let $\{ U_i \}_{i\in I}$ be a good cover of $M$ and let $U_{ij\dots
k} = U_{i} \cap U_{j} \cap \dots \cap U_{k}$.
The trivial bundle has a sections $s_i$ which are related by
$$
s_i = s_j [u_{ji}]
$$
where $[u_{ji}] \colon U_{ij} \to PU(\cH)$ for some $u_{ji} \colon
U_{ij}\to U(\cH)$ where
$u_{ij}u_{jk}u_{ki} = g_{ijk}1$ where $1$ is the
identity operator and the $g_{ijk}$ are non-zero scalars.

Over each of the $s_i(U_i)$  are sections $\sigma_i $ of the $\UK$
bundle $R$. We can compare
$\sigma_i$ and $\sigma_j[u_{ji}]$ so that
$$
\sigma_i = \sigma_j [u_{ji}] g_{ji}
$$
where $g_{ij} \colon U_{ij} \to \UK$. These satisfy
\begin{equation}
\label{eq:twisted}
g_{ki} = ([u_{ji}^{-1}] g_{kj} [u_{ji}]) g_{ji}.
\end{equation}

If $Y_i = \pi^{-1}(U_i)$ you can define a section of $R$ over all of $Y_i$ by
$\hat\sigma_i ( s_i [u]) = \sigma_i[u]$.  The transition functions for these
are $\hat g_{ij}$ where $\hat g_{ij} (s_j[u]) = [u^{-1}] g_{ji} [u]$ and the
identity \eqref{eq:twisted}  is equivalent to $\hat g_{ki} = \hat
g_{kj} \hat g_{ji}$.

\section{Examples}
\label{sec:examples}

This section contains calculations of twisted $K$-theory, mainly for
3 dimensional manifolds. In the ensuing computations, we sometimes make
use of the  following observation in defining the connecting
homomorphisms. Dixmier-Douady classes can be regarded canonically as
elements in
$K^1$-theory in 3 dimensions. This is because when $X$ is a
3 dimensional manifold, then it is a standard fact, since $SU(2) \cong
S^3$, that
$ H^3(X, \mathbb Z) = [X, SU(2)]$, where
$[\ ,\ ]$ means homotopy classes.
But $SU(2)$ includes canonically as a subgroup of
$U(\infty) = \displaystyle\lim_{\stackrel{\to}{n}} U(n)$, so that any
map $f: X \to
SU(2)$ can be regarded canonically as an element in $K^1(X) = [X,
U(\infty)]$. It follows that the map
$\phi: H^3(X, \mathbb Z) \to K^1(X)$ is a homomorphism of groups, for
3 dimensional
manifolds $X$. There is also a homomorphism $\Ch_3: K^1(X)\to H^3( X,
\mathbb Z)$
that is derived from the Chern character and is given by the formula
$\Ch_3(t) = \frac{1}{12\pi}\Tr((t^{-1}dt)^3)$. Noting that
$H^3(X, \mathbb Z)$ is torsion-free, a calculation shows that
the composition $\Ch_3\circ \phi$ is the identity, since
$\frac{1}{12\pi}\Tr((g^{-1}dg)^3)$ is the volume form of $SU(2)$
and so the differential form representative of a map
$f: X \to SU(2)$ is just $\Ch_3(f)$.
Therefore $\phi$
is injective, which allows us to identify Dixmier-Douady classes
with elements in $K^1(X)$.

% ------------------------------------------------------------------------
\subsection{The three-sphere}

We first discuss a few methods to compute the $K$-theory $K^\bullet(S^3)$
in the untwisted case, and then generalize to the twisted case.

% ------------------------------------------------------------------------
\subsubsection{Mayer-Vietoris}

Suppose $X=U_1\cup U_2$, where $U_i, i=1,2$, are closed subsets
of a locally compact space $X$. Then we have the short exact
sequence of $C^*$-algebras
\begin{equation}
\begin{CD}
0 @>>> C_0(X) @>\imath>> C_0(U_1) \oplus C_0(U_2) @>\pi>>
C_0(U_1\cap U_2) @>>> 0
\end{CD}
\end{equation}
and the associated six-term exact
(Mayer-Vietoris) sequence on $K$-theory \cite[Th. 4.18]{Kar} %
\footnote{There exists an analogous sequence if the $U_i$'s are
open subsets \cite[Th. 4.19]{Kar}.}
\begin{equation} \label{eqab}
\begin{CD}
K^0(X) @>\imath_*>> K^0(U_1)\oplus K^0(U_2) @>\pi_*>>
        K^0(U_1\cap U_2) \\
@AAA      && @VVV \\
K^1(U_1\cap U_2) @<\pi_*<< K^1(U_1)\oplus K^1(U_2) @<\imath_*<<
         K^1(X)
\end{CD}
\end{equation}
Now consider $X=S^3$.  Take for the $U_i$ the upper and lower
(closed) hemispheres $D_\pm$, respectively.  Then, since $D_\pm$
is contractible, we have $K^0(D_\pm)=\ZZ$, $K^1(D_\pm)=0$, while
$D_+\cap D_- \sim_h S^2$.  Hence \eqref{eqab} reduces to
\begin{equation}
\begin{CD}
K^0(S^3) @>\imath_*>> \ZZ \oplus \ZZ @>\pi_*>> K^0(S^2) \\
@AAA      && @VVV \\
K^1(S^2) @<\pi_*<< 0 @<\imath_*<<    K^1(S^3)
\end{CD}
\end{equation}
If we use the fact that $K^0(S^2)=\ZZ \oplus \ZZ$ and $K^0(S^2)=0$,
then we have a short exact sequence
\begin{equation}
\begin{CD}
0 @>>> K^0(S^3) @>\imath_*>> \ZZ \oplus \ZZ @>\pi_*>> \ZZ \oplus \ZZ
@>>> K^1(S^3) @>>> 0
\end{CD}
\end{equation}
where the map $\pi_*$ is easily seen to be given by $\pi_*(m,n)=(m-n,0)$.
We conclude $K^0(S^3)=\ZZ$, $K^1(S^3)=\ZZ$.

We can use the same procedure to compute the twisted $K$-theory of
$S^3$ (or, more generally, when
the $PU(\cH)$ bundle $\cE_{[H]}$ is trivial over $U_i$).
In that case, $C_0(X,\cE_{[H]})$ is given by pasting
$C_0(U_1)\otimes \cK$ and $C_0(U_2)\otimes \cK$ over
$U_1\cap U_2$ via a map $L_{[H]}\,:\,U_1\cap U_2\to PU(\cH)$, i.e.\
$$
C_0(X,\cE_{[H]})  = \{ (f_1,f_2)\ | \ f_i\in C_0(U_i)\otimes \cK\,,\
       f_1{}_{\lvert U_1\cap U_2} = L_{[H]} (f_2{}_{\lvert U_1\cap U_2} )
        \}  \,,
$$
and we have a short exact sequence
$$
\begin{CD}
0 @>>> C_0(X,\cE_{[H]}) @>\imath>> \bigoplus_i C_0(U_i)\otimes \cK
@>\pi>> C_0(U_1\cap U_2) \otimes \cK @>>> 0
\end{CD}
$$
where
$$
\imath(f_1,f_2) = f_1\oplus f_2 \,,\qquad
\pi(f_1\oplus f_2) = f_1{}_{\lvert U_1\cap U_2} -
        L_{[H]} (f_2{}_{\lvert U_1\cap U_2} ) \,.
$$
The associated six-term exact sequence in twisted $K$-theory
is given by \cite{Rosa}
\begin{equation}
\begin{CD}
K^0(X,[H]) @>\imath_*>> K^0(U_1)\oplus K^0(U_2) @>\pi_*>>
        K^0(U_1\cap U_2) \\
@AAA      && @VVV \\
K^1(U_1\cap U_2) @<\pi_*<< K^1(U_1)\oplus K^1(U_2) @<\imath_*<<
         K^1(X,[H])
\end{CD}
\end{equation}
and in the case of $S^3$ collapses to
\begin{equation}
\begin{CD}
0 @>>> K^0(S^3,[H]) @>\imath_*>> \ZZ \oplus \ZZ @>\pi_*>> \ZZ \oplus \ZZ
@>>> K^1(S^3,[H]) @>>> 0
\end{CD}
\end{equation}
where now the map $\pi_*$ is given by $\pi_*(m,n)=(m-n,-nN)$
if $[H]=N[H_0]$ where $[H_0]$ is the generator of $H^3(S^3,\ZZ)=\ZZ$.
We conclude $K^0(S^3,[H])=0$, $K^1(S^3,[H])=\ZZ/N\ZZ$.
This computation was initially performed in \cite{Rosa} and was
recently reviewed in the context of D-branes in \cite{Maly}.

% ------------------------------------------------------------------------
\subsubsection{The three-sphere at infinity}

We can think of $S^3$ as being the boundary of the closed four-ball
$B^4$.  This leads to a short exact sequence
\begin{equation} \label{eqba}
\begin{CD}
0 @>>> C_0(\RR^4) @>\imath>> C(B^4) @>>> C(S^3) @>>> 0
\end{CD}
\end{equation}
and associated six-terms sequence in $K$-theory
\begin{equation}
\begin{CD}
K^0(\RR^4) @>\imath_*>> K^0(B^4) @>>>   K^0(S^3) \\
        @AAA      &            &      @VVV \\
K^1(S^3)   @<<< K^1(B^4) @<<<    K^1(\RR^4)
\end{CD}
\end{equation}
Using Bott-periodicity, $K^0(\RR^4)=\ZZ$, $K^1(\RR^4)=0$, while
contractibility of $B^4$ implies $K^0(B^4)=\ZZ$, $K^1(B^4)=0$.
Using furthermore that the map $\imath_*:K^0(\RR^4)\to K^0(B^4)$
is trivial in this case, we again have $K^0(S^3)=K^1(S^3)=\ZZ$.
However, since in general a $PU(\cH)$-bundle $\cE_{[H]}$ over
$S^3$ does not extend to $B^4$ we can not generalize \eqref{eqba}
to the twisted case.

% ------------------------------------------------------------------------
\subsubsection{Excision of a point}

Excising a point $x_0$ from $S^3$ we have
\begin{equation} \label{eqca}
\begin{CD}
0 @>>> C_0(\RR^3) @>>> C(S^3) @>>> C(\{x_0\}) @>>> 0
\end{CD}
\end{equation}
and, accordingly,
\begin{equation} \label{eqcb}
\begin{CD}
K^0(\RR^3) @>>> K^0(S^3) @>>>   K^0(\{x_0\}) \\
        @AAA      &            &      @VV\Delta V \\
K^1(\{x_0\})   @<<< K^1(S^3) @<<<    K^1(\RR^3)
\end{CD}
\end{equation}
Again, using Bott-periodicity, $K^0(\RR^3)=0$, $K^1(\RR^3)=\ZZ$, while
$K^0(\{x_0\}) =\ZZ$, $K^1(\{x_0\}) =0$.  In this case the
connecting map $\Delta:K^0(\{x_0\})\to K^1(\RR^3)$ is trivial.
And thus, we conclude $K^0(S^3)=K^1(S^3)=\ZZ$.
In this case we can generalize \eqref{eqca} to the twisted case, namely
\begin{equation} \label{eqcc}
\begin{CD}
0 @>>> C_0(\RR^3)\otimes \cK @>>> C(S^3,\cE_{[H]}) @>>>
C(\{x_0\})\otimes \cK  @>>> 0
\end{CD}
\end{equation}
The associated six-term sequence in twisted $K$-theory is similar
to \eqref{eqcb}, except that $\Delta:K^0(\{x_0\})=\ZZ\to K^1(\RR^3)=\ZZ$,
is now given by $\Delta(m) = mN$.
We conclude again $K^0(S^3,[H])=0$, $K^1(S^3,[H])=\ZZ/N\ZZ$.

% ------------------------------------------------------------------------
\subsection{Product of one- and two-sphere}

The case $X=S^1\times S^2$ is interesting since an explicit
realization of the principle $PU(\cH)$-bundles over $X$,
for $[H]\in H^3(S^1\times S^2,\ZZ)=\ZZ$, is known
\cite{Bry}.
We can compute $K^\bullet(S^1\times S^2,[H])$ using the Mayer-Vietoris
sequence as in Section 8.1.1, but the analogue of the procedure in
section 8.1.3 is more convenient.  Take a point $x_0\in S^1$. We have
\begin{equation} \label{eqda}
\begin{CD}
0 @>>> C_0(\RR\times S^2) @>>> C(S^1\times S^2) @>>>
C(\{x_0\}\times S^2) @>>> 0
\end{CD}
\end{equation}
Using $K^n(\RR\times S^2) = K^{n+1}(S^2)$ and
$K^\bullet(\{x_0\}\times S^2) =
K^\bullet(S^2)$, we have for the twisted analogue of \eqref{eqda}
\begin{equation} \label{eqdb}
\begin{CD}
K^1(S^2) @>>> K^0(S^1\times S^2,[H]) @>>>   K^0(S^2) \\
        @AAA      &            &      @VV\Delta V \\
K^1(S^2)   @<<< K^1(S^1\times S^2,[H]) @<<<    K^0(S^2)
\end{CD}
\end{equation}
Now, $K^0(S^2) = \ZZ\oplus \ZZ$ and $K^1(S^2)=0$.  The connecting
map $\Delta: K^0(S^2) \to K^0(S^2)$ corresponds to taking the cup-product
with $[H]$.  Say, if $[\omega]$ is the generating line-bundle of $K^0(S^2)$
then $\Delta (m\cdot 1 + n [\omega]) = (m\cdot 1 + n [\omega])\cup [H]$.
We conclude $\Delta: (m,n) = (0,mN)$.  Hence,
$K^0(S^1\times S^2,[H]) = \ZZ$ and $K^1(S^1\times S^2,[H]) = \ZZ \oplus
(\ZZ/N\ZZ)$ for $N\neq0$, while for $N=0$ we have
$K^0(S^1\times S^2,[H]) = K^1(S^1\times S^2,[H]) = \ZZ \oplus\ZZ$,
as it should.
The same conclusion is reached from the (twisted) Atiyah-Hirzebruch
spectral sequence, cf.\ \cite{Rosa}.

% ------------------------------------------------------------------------
\subsection{The real projective three-space}

The $K$-theory for the real projective spaces $\RP^n$ is given in
\cite[Prop. 2.7.7]{Ati}.
\begin{align}
\widetilde{K}^0(\RP^{2n+1}) = \widetilde{K}^0(\RP^{2n}) & =\ZZ_{2^n}\,,
& \nonumber \\
K^1(\RP^{2n+1}) & = \ZZ \,, & K^1(\RP^{2n}) & = 0 \,.
\end{align}
For $\RP^3$ part of this result is derived by looking
at the six-term sequence related to the short exact sequence
\begin{equation} \label{eqea}
\begin{CD}
0 @>>> C_0(\RR^3) @>>> C(\RP^3) @>>> C(\RP^2) @>>> 0
\end{CD}
\end{equation}
i.e., the exact sequence corresponding to the pair $(\RP^3,\RP^2)$.
The associated six-term exact sequence is
\begin{equation} \label{eqeb}
\begin{CD}
K^0(\RR^3) @>>> K^0(\RP^3) @>>>   K^0(\RP^2) \\
        @AAA      &            &      @VV\Delta V \\
K^1(\RP^2)   @<<< K^1(\RP^3) @<<<    K^1(\RR^3)
\end{CD}
\end{equation}
Now, $K^0(\RR^3)=0$ and $K^1(\RR^3)=\ZZ$.  So, the result of \cite{Ati},
$K^0(\RP^3)=K^0(\RP^2) = \ZZ \oplus \ZZ_2$ and $K^1(\RP^2)=0$,
$K^1(\RP^3)=\ZZ$
is perfectly consistent with \eqref{eqea} provided the connecting map
$\Delta : K^0(\RP^2) \to K^1(\RR^3)$ vanishes in this case (which is not
too hard to check independently, i.e.\ as in \eqref{eqdb}).
In the twisted version of \eqref{eqea} and \eqref{eqeb} the connecting
map $\Delta$ is given by $\Delta (m,n) = mN$, if $[H]$
is $N$ times the generator of $H^3(\RP^3,\ZZ)=\ZZ$.
Thus $K^0(\RP^3,[H]) = \ZZ_2$ and $K^1(\RP^3,[H]) = \ZZ/N\ZZ$.

Alternatively, we may use $\RP^3=S^3/\ZZ_2$ and
compute $K^\bullet(\RP^3)$ through the $\ZZ_2$-equivariant $K$-theory
of $S^3$.  In that case, however, none of the sequences for $S^3$
discussed in Section 1 are appropriate since we need $\ZZ_2$ to act
on the subspace.  We can however slightly modify \eqref{eqca} by
cutting out two points $\{x_0,x_1\} \in S^3$ related by the $\ZZ_2$ action.
I.e.\
\begin{equation} \label{eqec}
\begin{CD}
0 @>>> C_0(\RR \times S^2) @>>> C(S^3) @>>> C(\{x_0,x_1\}) @>>> 0
\end{CD}
\end{equation}
The associated six-term sequence in $\ZZ_2$-equivariant $K$-theory
is
\begin{equation} \label{eqed}
\begin{CD}
K^0_{\ZZ_2}(\RR\times S^2) @>>> K^0(\RP^3) @>>>   K^0(\{x_0\}) \\
        @AAA        &               &                      @VV\Delta V \\
K^1(\{x_0\})   @<<< K^1(\RP^3) @<<<    K^1_{\ZZ_2}(\RR\times S^2)
\end{CD}
\end{equation}
where we have used $K^\bullet_{\ZZ_2}(\{x_0,x_1\}) =
K^\bullet(\{x_0\})$ since
$\ZZ_2$ acts freely on $\{x_0,x_1\}$.  Note however that while $\ZZ_2$
does act freely on $\RR\times S^2$ it does not act freely on $\RR$
separately.  Hence it would be wrong to conclude that
$K^0_{\ZZ_2}(\RR\times S^2)$ is equal to
$K^0(\RR\times \RP^2) = K^1(\RP^2) = 0$.  In fact, the connecting map
$\Delta$ vanishes in this case. Using our result for $K^\bullet(\RP^3)$
then yields
$K^0_{\ZZ_2}(\RR\times S^2) = \ZZ_2$, $K^1_{\ZZ_2}(\RR\times S^2) = \ZZ$
(this also follows from \cite[Prop. 2.4]{Kar}).  In the twisted case
$\Delta (m) = mN$ and again we find
$K^0(\RP^3,[H]) = \ZZ_2$ and $K^1(\RP^3,[H]) = \ZZ/N\ZZ$.

% ------------------------------------------------------------------------
\subsection{Lens spaces}

In the case of a Lens space $L_p=S^3/\ZZ_p$, for $p$ a prime, we can
compute $K^\bullet(L_p)$ as in \cite{Ati} with the result
\begin{equation}
K^0(L_p) = \ZZ\oplus \ZZ_p\,,\qquad K^1(L_p)= \ZZ \,.
\end{equation}
Generalizing the equivariant computation of section 3 immediately gives
\begin{equation}
K^0(L_p,[H]) = \ZZ_p\,,\qquad K^1(L_p,[H])= \ZZ/N\ZZ \,.
\end{equation}
D-branes on Lens spaces were recently considered in \cite{MMS}.

% ------------------------------------------------------------------------
\subsection{Group manifolds}

In the case of $G=SU(n)$ we have \\
$H^\bullet(G,\QQ) = \bigwedge \{ c_3,c_5,\ldots,c_{2n-1} \}$
where $c_n\in H^n(G,\QQ)$.  The $K$-groups are given similarly in
terms of certain appropriately normalized linear combinations of the
$c_n$.  Since the generator of $H^3(G,\ZZ)=\ZZ$ corresponds to
$c_3$  (appropriately normalized), the third differential $d_3$ in the
(twisted) Atiyah-Hirzebruch spectral sequence corresponds to taking the
wedge product with $Nc_3$ if $[H]=Nc_3$.
Thus
\begin{equation}
{\rm Ker} \ d_3 = \ZZ c_3 \wedge \left( \bigwedge \{ c_5,\ldots,c_{2n-1} \}
       \right)
\end{equation}
while
\begin{equation}
{\rm Im} \ d_3 = N \ZZ c_3 \wedge \left( \bigwedge \{ c_5,\ldots,c_{2n-1} \}
       \right) \,.
\end{equation}
I.e., at the 3rd term of the spectral sequence, we have
\begin{equation}
E_3^\bullet(SU(n),[H]) = (\ZZ/N\ZZ) \ c_3 \wedge \left(
\bigwedge \{ c_5,\ldots,c_{2n-1} \} \right) \,.
\end{equation}
The spectral sequence collapses at the 3rd term for $G=SU(2)$, e.g.
\begin{equation}
K^0(SU(2),[H])  = 0\,, \qquad  K^1(SU(2),[H])  = \ZZ/N\ZZ\,, 
\end{equation}
(in agreement with the results of Section 8.1.1), but for 
$G=SU(n), n>2$, the higher order differentials are nonzero
\cite{Hop} (see also \cite{MalMooSei}).
D-branes on group manifolds were studied in, e.g.,
\cite{AlSc,FigSta,FS}.

%---------------------------------------------------------------------------- 
\section{The Chern character in the non-torsion case} 

\subsection{$U_{\mathcal{K}}$ bundle gerbe modules} 

Recall that in the finite dimensional case we were able 
to consider connections $\nabla$ on a finite rank bundle gerbe module 
$E$ which were compatible with a bundle gerbe connection 
on the bundle gerbe $L$, so-called `bundle gerbe 
module connections'.  If $F_{\nabla}$ denoted the 
curvature $2$-form of $\nabla$, we were able to show 
that the $2$-form $F_{\nabla} + fI$ with values in the 
bundle $\Omega^2(\End(E))$ was compatible with the descent isomorphism 
for $\End(E)$, where $f$ was the `curving' for the 
bundle gerbe connection on $L$.  It followed that the 
forms $\tr(F_{\nabla} + fI)^k$ descend to $M$.   

In the case of infinite rank bundle gerbe modules, to make 
sense of the trace, we need to restrict our attention to 
bundle gerbe modules with a reduction of the structure 
group to $U_1$, the group of unitaries differing from the 
identity by a trace class operator.  As remarked earlier, 
in Note~\ref{note:replace U_K with U_1}, 
in the definition of twisted $K$-theory 
we can replace Fred by any homotopy equivalent space, as 
long as that space is $PU(\mathcal{H})$ equivariant.  
The notion of a bundle 
gerbe module connection continues to make sense in this 
setting however it is not possible to find module connections 
so that the bundle-valued $2$-forms $F_{\nabla} + fI$ 
take trace class values, ie lie in the adjoint bundle 
$\Omega^2 (\ad(P))$ associated to the $PU(\mathcal{H})$ covariant 
principal $U_1$ bundle $P$ via the adjoint action of 
$U_1$ on its Lie algebra $\mathcal{L}_1$, the ideal 
of trace class operators on $\mathcal{H}$.  Instead, 
given a pair of $PU(\mathcal{H})$ covariant principal 
$U_1$ bundles $P$ and $Q$, defining a class in 
twisted $K$-theory, we can consider differences of 
bundle-valued $2$-forms $(\mathcal{F}_P + fI) - (\mathcal{F}_Q 
+fI)$ coming from module connections on the  
Hilbert vector bundles associated to $P$ and $Q$.  We will 
show that it is still possible to make sense of the 
trace in this setting and that we can define 
forms on $M$ representing classes in the 
twisted cohomology group $H^{\bullet}(M;H)$.  We propose that 
these forms on $M$ define the Chern character for reduced twisted 
$K$-theory $\tilde{K}^0(M;[H])$.  Recall that the Chern character for 
(reduced) twisted $K$-theory is a homomorphism $ch_{[H]}\colon 
\tilde{K}^0(M;[H])\to H^\bullet(M;[H])$.  $ch_{[H]}$ is 
uniquely characterised by requiring that it is functorial 
with respect to pullbacks, respects the 
$\tilde{K}^0(M)$-module structure of $\tilde{K}^0(M;[H])$ 
and reduces to the ordinary Chern character in the 
untwisted case when $[H] = 0$.

\subsection{Remarks on the projective unitary group} 

Dixmier and Douady's 1963 work on continuous fields of 
$C^*$-algebras exploited the fact that there is a 
natural bijection between $H^3(M;\ZZ)$ and isomorphism 
classes of principal $PU$ bundles on $M$.  They used 
the strong operator topology on $U(\mathcal{H})$, the group of unitary 
operators on an infinite dimensional separable Hilbert space 
$\mathcal{H}$.  Neither $U(\mathcal{H})$ with the 
strong operator topology nor $PU(\mathcal{H})$ with the 
induced topology are Lie groups.     
In 1965 Kuiper proved that $U(\mathcal{H})$ 
equipped with the norm topology is contractible.  
$U(\mathcal{H})$, equipped with the norm topology, is a Lie group 
(see for instance \cite{Milnor}) 
and one can show that $PU(\mathcal{H})$ equipped with the 
topology induced by the norm topology on $U(\mathcal{H})$ 
is a Lie group modelled locally on the quotient 
$Lie(U(\mathcal{H}))/i\RR$ \cite{Tol}.

\subsection{Twisted cohomology} 

There are several definitions of twisted cohomology 
that are well known among experts and which are all 
probably equivalent.  One such definition is 
given by Atiyah in \cite{Ati2}.  We give another 
definition here.   
If $H$ is a closed, differential $3$-form on 
$M$ then we can use $H$ to introduce 
a `twist' on the usual cohomology of $M$ and consider the 
twisted cohomology group $H^\bullet(M;H)$.   
$H^\bullet(M;H)$ is constructed 
from the algebra $\Omega^\bullet(M)$ of differential forms 
on $M$ by introducing a twisted differential $\delta$ on $\Omega^\bullet(M)$ 
given by $\delta = d-H$, where $d$ is the usual exterior derivative of 
differential forms on $M$.  It is easy to see that 
$\delta^2 = 0$ using the fact that $H$ is of degree 
three and hence $H^2 = 0$.  We then set 
$$ 
H^\bullet(M;H) = \text{ker}\{\delta\colon \Omega^\bullet(M) 
\to \Omega^\bullet(M)\}/\text{im}\{\delta\colon \Omega^\bullet(M) 
\to \Omega^\bullet(M)\}. 
$$
$H^\bullet(M;H)$ is then a group under addition which 
satisfies the obvious functorial property that a 
smooth map $f\colon N\to M$ induces a homomorphism 
$f^*\colon H^\bullet(M;H) \to H^\bullet(N;f^*H)$.   
Note that although there is no algebra structure on 
$H^\bullet(M;H)$, there exist homomorphisms 
$H^\bullet(M;H)\otimes H^\bullet(M;H') \to 
H^\bullet(M;H+H')$.  When $H = H' = 0$ this is 
just the usual wedge product of forms.  If $H' = 0$ only 
then this map defines an action of the ordinary cohomology algebra  
$H^\bullet(M)$ on $H^\bullet(M;H)$ making 
$H^\bullet(M;H)$ into 
a $H^\bullet(M)$-module.  Note that if $\lambda$ is a 
$2$-form on $M$ then we can define a map $H^\bullet(M;H) 
\to H^\bullet(M;H+d\lambda)$ by sending a representative 
$\omega$ of a class $[\omega]$ in $H^\bullet(M;H)$ to 
the class in $H^\bullet(M;H+d\lambda)$ represented by 
$\exp(\lambda)\omega$.  This is well defined, since if 
$\omega' = \omega + d\mu - H\mu$ then 
\begin{eqnarray*} 
\exp(\lambda)\omega' & = & \exp(\lambda)\omega   
+ \exp(\lambda)d\mu - H\exp(\lambda)\mu           \\ 
& = & \exp(\lambda)\omega + d(\exp(\lambda)\mu) - 
(H+ d\lambda)\exp(\lambda)\mu. 
\end{eqnarray*} 
Suppose that the closed $3$-form 
$H$ is the representative for the image, in real 
cohomology, of the Dixmier-Douady class of a principal 
$PU(\mathcal{H})$ bundle $Y$ on $M$.  We interpret 
$H$ as the $3$-curvature of a bundle gerbe connection 
$\nabla_L$ and curving $f$ for the lifting bundle gerbe 
$L\to Y^{[2]}$ associated to $Y$.  Recall that a curving 
$f$ for $\nabla_L$ satisfies $\delta(f) = F_{\nabla_L}$ 
and $df = \pi^*H$ where $F_{\nabla_L}$ is the curvature of the 
connection $\nabla_L$ on the line bundle $L$ and $\pi\colon 
Y\to M$ is the projection.  Any two curvings for 
$\nabla_L$ differ by the pullback of a $2$-form $\lambda$ 
on $M$.  We define $H^\bullet(M;[H])$ to be the set 
of equivalence classes of quadruples $([\omega],L,\nabla_L,f)$, 
where $L$ is the lifting bundle gerbe for a 
principal $PU(\cH)$ bundle $Y$ with Dixmier-Douady 
class $[H]$, 
$f$ is a curving for the bundle gerbe 
connection $\nabla_L$ with $3$-curvature 
$H$ (so that $df = \pi^*H$) and $[\omega] \in H^\bullet(M;H)$.  
If $J$ is the lifting bundle gerbe for another principal 
$PU(\cH)$ bundle $X$ on $M$ whose Dixmier-Douady class 
is also equal to $[H]$, then $Y$ and $X$ are isomorphic.  
This extends to an isomorphism between the lifting 
bundle gerbes $L$ and $J$.  We declare 
two quadruples $([\omega],L,\nabla_L,f_L)$ 
and $([\omega'],J,\nabla_J,f_J)$ 
to be equivalent if, under the ismorphism 
$L = J$, we have $\nabla_J = \nabla_L + \delta(\rho)$ 
for some complex valued $1$-form $\rho$ on $X$,   
$f_J = f_L + d\rho + \pi^*\lambda$ for some 
$2$-form $\lambda$ on $M$  
and $[\omega'] = [\exp(\lambda)\omega]$, where 
$[\omega] \in H^\bullet(M;H)$ and $[\omega'] \in 
H^\bullet(M;H+d\lambda)$.  Here we are identifying 
$[\omega']$ with the image of $[\omega]$ under the isomorphism 
of complexes $H^\bullet(M;H) \to H^\bullet(M;H+d\lambda)$ 
defined above.  For a curving $f$ for 
$\nabla_L$ we will also define $H^\bullet(M;L,\nabla_L,f)$ to 
be $H^\bullet(M;H)$ where $H$ is the $3$-curvature 
of the pair $(\nabla_L,f)$.  Then $H^\bullet(M;[H])$ 
is equal to the quotient of the union of the $H^\bullet(M;L,\nabla_L,f)$ 
over all bundle gerbe connections $\nabla_L$ and curvings $f$ 
on the lifting bundle gerbes $L$,    
under the equivalence relation defined above.  The twisted cohomology 
groups satisfy the following properties.  If 
$f\colon N\to M$ is a smooth map then there is an 
induced map $f^*\colon H^\bullet(M;[H]) \to 
H^\bullet(N;f^*[H])$.  $H^\bullet(M;[H])$ is a module 
over $H^\bullet(M)$.  This in turn follows from the 
property that if $[H]$ and $[H']$ are classes in 
$H^3(M;\mathbb{Z})$, then there is a homomorphism 
$H^\bullet(M;[H])\otimes H^\bullet(M;[H'])\to 
H^\bullet(M;[H]+[H'])$.  These properties are 
analogous to those for twisted $K$-theory.    

\subsection{Defining the Chern Character} 

Suppose $L\to Y^{[2]}$ is a bundle gerbe with 
bundle gerbe connection $\nabla_L$.  
Recall that a module connection $\nabla_E$ 
on a bundle gerbe module $E$ for $L$ is a connection on 
the vector bundle $E$ which is compatible with the 
bundle gerbe connection $\nabla_L$, ie under the 
isomorphism $\pi_1^{-1}E\otimes L\to \pi_2^{-1}E$ 
the tensor product connection $\pi_1^{-1}\nabla_E\otimes 
\nabla_L$ on $\pi_1^{-1}E\otimes L$ is mapped into 
the connection $\pi_2^{-1}\nabla_E$ on $\pi_2^{-1}E$.  
Suppose now that $L\to Y^{[2]}$ is the lifting bundle 
gerbe for the principal $PU(\mathcal{H})$ bundle 
$Y\to M$ and that $\nabla_L$ is a bundle gerbe 
connection on $L$ with curving $f$ such that the 
associated $3$-curvature (which represents the image, in 
real cohomology, of the Dixmier-Douady class of $L$) is 
equal to the closed, integral $3$-form $H$.  If $E$ is a 
$U_1$ bundle gerbe module for $E$ then we can consider 
module connections $\nabla_E$ on $E$; however, as remarked 
above, the algebra valued $2$-form $F_E +fI$ cannot take 
trace class values (here $F_E$ denotes the curvature 
of the connection $\nabla_E$).  If $F$ is another 
$U_1$ bundle gerbe module for $L$, so that the difference 
$E-F$ represents a class in $\tilde{K}^0(M;[H])$ 
under the isomorphism of Proposition~\ref{prop:U_K 
bg modules equals K(M;H)}, we can consider  
module connections $\nabla_E$ and $\nabla_F$ on $E$ and 
$F$ respectively such that the difference of connections 
$\nabla_E - \nabla_F$ is trace class.  By this we mean 
that in local trivialisations of $E$ and $F$ such that 
the connections $\nabla_E$ and $\nabla_F$ are given by 
$d + A_E$ and $d + A_F$ respectively, the difference 
$A_E - A_F$ is trace class.  It follows that the difference   
of curvatures $F_E - F_F$ in local trivialisations of $E$ and $F$ 
respectively is trace class.  One can show that it is 
always possible to find such module connections.    

It follows that the differences $(F_E + fI) - (F_F +fI)$ 
and hence $(F_E +fI)^k - (F_F +fI)^k$ take trace class 
values (considered in local trivialisations of $E$ 
and $F$ respectively).  Therefore the $2k$-forms 
$\tr((F_E +fI)^k - (F_F +fI)^k)$ on $Y$ are well defined.   
To see this, note that in a local trivialisation of $E$, $F_E$ is given
by operator valued $2$-forms $F_E^i$ which are related 
on overlaps by $F_E^j = g_{ij}^{-1}F_E^i g_{ij}$, where 
$g_{ij}$ are the $U_1$ valued transition functions for 
$E$.  Similarly, in local trivialisations of $F$ defined 
over the same open cover of $Y$ as the local trivialisations 
of $E$, the curvature $2$-form $F_F$ is given locally 
by the operator valued $2$-forms $F_F^i$ which are related 
on overlaps by $F_F^j = h_{ij}^{-1}F_F^ih_{ij}$ where $h_{ij}$ are
the $U_1$ valued transition functions for $F$.  Therefore 
the local $2k$-forms $\tr((F_E^i + fI)^k - (F_F^i + fI)^k)$ 
define global forms, since    
\begin{eqnarray*} 
&   & \tr((F_E^j +fI)^k - (F_F^j +fI)^k) \\ 
& = & \tr(g_{ij}^{-1}(F_E^i +fI)^k g_{ij} - 
h_{ij}^{-1}(F_F^i +fI)^k h_{ij})                \\ 
& = & \tr(g_{ij}^{-1}(F_E^i +fI)^k 
g_{ij} - (F_E^i +fI)^k) 
+ \tr((F_E^i +fI)^k - (F_F^i +fI)^k)  \\  
&   & + \tr((F_F^i +fI)^k - h_{ij}^{-1}
(F_F^i +fI)^k h_{ij}) \\ 
& = & \tr((F_E^i +fI)^k - (F_F^i +fI)^k). 
\end{eqnarray*} 
We want to know that the forms we have defined live on 
$M$.  This follows from the fact that the $F_E$ 
and $F_F$ are curvatures of module connections, and 
therefore satisfy equation~\ref{eq:module connection transfm law}.  
More precisely, suppose that $\mathcal{U} = \{U_{\alpha}\}_{\alpha\in 
\Sigma}$ is an open cover of $M$ such that there exist 
local sections $s_{\alpha}\colon U_{\alpha}\to Y$ 
of the $PU(\mathcal{H})$ bundle $Y\to M$.  Suppose that 
$Y$ has transition functions $g_{\alpha\beta}$ relative 
to this open covering.  If $E$ is a $U_1$ bundle gerbe module 
for $L$ then we can use the sections $s_{\alpha}$ to pullback 
$E$ to form Hilbert vector bundles $E_{\alpha}$ on 
$U_{\alpha}$ whose structure group reduces to $U_1$.  
The transition functions $g_{\alpha\beta}$ for $Y$  
then provide maps $E_{\alpha}\stackrel{g_{\alpha\beta}}{\to} 
E_{\beta}$.  If $F$ is the curvature of a module 
connection on $E$, then the pullbacks $F_{\alpha} 
+ f_{\alpha}I$ are related by $F_{\beta} + 
f_{\beta}I = \hat{g}_{\alpha\beta}^{-1}(F_{\alpha} 
+ f_{\alpha}I)\hat{g}_{\alpha\beta}$.  On taking traces 
of differences of powers as above we see that these forms 
are globally defined.    
Note that 
\begin{equation} 
\label{eq:character form} 
\tr(\exp(F_E + fI) - \exp(F_F + fI)) 
= \exp(f)\tr(\exp(F_E) - \exp(F_F)). 
\end{equation} 
We have the following Proposition. 
\begin{proposition} 
Suppose that $E$ and $F$ are $U_1$ bundle gerbe modules for 
the lifting bundle gerbe $L\to Y^{[2]}$ equipped with a 
bundle gerbe connection $\nabla_L$ and curving $f$, such 
that the associated $3$-curvature is $H$.  
Suppose that $\nabla_E$ and $\nabla_F$ are module 
connections on $E$ and $F$ respectively 
such that the difference $\nabla_E - \nabla_F$ is trace class, 
considered in local trivialisations of $E$ and $F$.    
Let $ch_H(\nabla_E,\nabla_F) \in \Omega^\bullet(M)$ denote the differential 
form on $M$ whose lift to $Y$ is given by $\exp(f) \textnormal{tr} ( 
\exp(F_E) - \exp(F_F))$.  Then 
$ch_H(\nabla_E,\nabla_F)$ is closed with respect to the twisted differential 
$d-H$ on $\Omega^\bullet(M)$ and hence represents a class 
in $H^\bullet(M;H)$.  The class $[ch_H(\nabla_E,\nabla_F)]$ is independent 
of the choice of module connections $\nabla_E$ and 
$\nabla_F$ on $E$ and $F$.  
\end{proposition} 

\begin{remark} 
Some care needs to be taken when working with connections 
on infinite dimensional vector bundles, as it is not always clear 
that the difference of two connections on $E$ is a section of 
$\Omega^1(\End(E))$, where $\End(E)$ denotes the bundle on 
$Y$ whose fibre at $y$ is the space of all bounded linear 
operators $E_y \to E_y$.  We will avoid this problem by 
fixing a module connection on $E$ and then only consider module 
connections which differ from this fixed connection by 
a section of $\Omega^1(\End(E))$.    
\end{remark} 

To show that $ch_H(\nabla_E,\nabla_F)$ is closed under $d-H$ it is 
sufficient to show that $\tr(\exp(F_E) - 
\exp(F_F))$ is closed.  
We have   
\begin{eqnarray*} 
&   & d\tr((F_E^i)^k - (F_F^i)^k) \\ 
& = & \tr(\sum F_E^i \cdots dF_E^i 
\cdots F_E^i - \sum F_F^i \cdots 
dF_F^i \cdots F_F^i)              \\ 
& = & \tr(\sum F_E^i \cdots [F_E^i, 
A_E^i]\cdots F_E^i - \sum 
F_F^i \cdots [F_F^i,A_E^i] 
\cdots F_F^i)                                 \\  
& = & \tr(\sum F_E^i \cdots [F_E^i, 
A_E^i] \cdots F_E^i - \sum F_F^i 
\cdots [F_F^i,A_E^i]\cdots F_F^i \\  
&   & + \sum F_F^i \cdots [F_F^i,A_E^i 
- A_F^i]\cdots F_F^i)                    \\ 
& = & \tr(\sum F_E^i\cdots [F_E^i, 
A_E^i]\cdots F_E^i - \sum F_F^i 
\cdots [F_F^i,A_E^i]\cdots F_F^i) \\  
&    & + \tr(\sum F_F^i \cdots [F_F^i, 
A_E^i - A_F^i]\cdots F_F^i)           
\end{eqnarray*} 
Note that the second term makes sense as the difference of the 
two module connections $A_E^i - A_F^i$ is 
trace class.  The first term vanishes by the usual argument 
for Chern-Weil theory, the trace is invariant: $\tr(g^{-1}Ag) 
= \tr(A)$ for any $g \in U(\mathcal{H})$ so long as $A$ is trace class.  
The second term also vanishes: we could write it as 
$$
k\tr(F_F^i \cdots F_F^i[F_F^i, 
A_E^i -A_F^i]),  
$$
since the $F_F$ are all even forms we can shuffle 
them around in the trace to show that this is zero.  

We now show that the forms $ch_H(\nabla_E,\nabla_F)$ are independent of 
the choice of module connections $\nabla_E$ 
and $\nabla_F$ on the $U_1$ bundle gerbe modules $E$ and $F$.   
So suppose that  
$\nabla'_E$ and $\nabla'_F$ is another pair of module 
connections on the $U_1$ bundle gerbe modules $E$ and 
$F$ respectively such that the difference $\nabla'_E - 
\nabla'_F$ is trace class when considered in local trivialisations 
of $E$ and $F$.   
Form the families of module connections $\nabla_E(t)$ and 
$\nabla_F(t)$ on $E$ and $F$ respectively given 
by $\nabla_E(t) = t\nabla_E + (1-t)\nabla'_E$ 
and $\nabla_F(t) = t\nabla_F +(1-t)\nabla'_F$.  It is 
clear that the difference $\nabla_E(t) - \nabla_F(t)$ 
is trace class.   
Consider the $2k-1$ forms $\Psi_k$ on $Y$ given locally by 
\begin{equation} 
\Psi_k = \tr (\frac{d}{dt}(A_E(t))(F_E(t) 
+fI)^{k-1} - \frac{d}{dt}(A_F(t))(F_F(t) 
+fI)^{k-1}). 
\end{equation} 
It is easy to see that $\Psi_k$ is in fact a global 
$2k-1$-form on $Y$ and that moreover 
$\Psi_k$ descends to a form on $M$.  We 
calculate the exterior derivative of $\Psi_k$: 
\begin{eqnarray*} 
&   & d\Psi_k                        \\ 
& = & \tr(\frac{d}{dt}(dA_E(t))(F_E(t)  
+ fI)^{k-1} - \frac{d}{dt}(dA_F(t))(F_F(t) + fI)^{k-1} \\ 
&   & +\frac{d}{dt}(A_F(t))\sum (F_F(t) 
+ fI)\cdots ([F_F(t),A_F(t)] + HI)\cdots 
(F_F(t) + fI)                                      \\ 
&   & -\frac{d}{dt}(A_E(t))\sum (F_E(t) 
+ fI)\cdots ([F_E(t),A_E(t)] + HI) 
\cdots (F_E(t) + fI))                                \\ 
& = & \tr(\frac{d}{dt}(dA_E(t))(F_E(t) 
+ fI)^{k-1} - \frac{d}{dt}(dA_F(t))(F_F(t) 
+ fI)^{k-1}                                                   \\ 
&   & +\frac{d}{dt}(A_F(t))\sum (F_F(t) + fI) 
\cdots [F_F(t),A_F(t)]\cdots (F_F(t) + fI) \\ 
&   & - \frac{d}{dt}(A_E(t))\sum (F_E(t) 
+ fI)\cdots [F_E(t),A_E(t)]\cdots 
(F_E(t) + fI))                                    \\ 
&    & + (k-1)H\Psi_{k-1}.  
\end{eqnarray*} 
We now examine $d/dt$ of $\tr((F_E(t) + fI)^k 
- (F_F(t) + fI)^k)$.  Using $\dot{F}_t = d\dot{\theta}_t 
+ \dot{\theta}_t\theta_t + \theta_t\dot{\theta}_t$ we get 
\begin{eqnarray*} 
&   & \frac{d}{dt}\tr((F_E(t)+fI)^k - 
(F_F(t) + fI)^k)                         \\ 
& = & \tr(\sum (F_E(t) + fI) \cdots 
(\frac{d}{dt}(dA_E(t)) + \frac{d}{dt}(A_E(t)) 
A_E(t)                                    \\  
&   & + A_E(t)\frac{d}{dt}(A_E(t))) 
\cdots (F_E(t) + fI)                       
 - \sum (F_F(t)+fI)\cdots             \\ 
&   & (\frac{d}{dt} 
dA_F(t) + \frac{d}{dt}(A_F(t))A_F(t) 
+ A_F(t)\frac{d}{dt}(A_F(t)))\cdots 
(F_F(t) + fI))                                \\ 
& = & k\tr(\frac{d}{dt}(dA_E(t))(F_E(t) 
+ fI)^{k-1} - \frac{d}{dt}(dA_F(t))(F_F(t) 
+ fI)^{k-1})                                                 \\ 
&   & - k\tr(\frac{d}{dt}(A_E(t))\sum (F_E(t) + fI) 
\cdots [F_E(t),A_E(t)]\cdots 
(F_E(t) + fI)                                      \\ 
&   & + \frac{d}{dt}(A_F(t))\sum (F_F(t) + fI) 
\cdots [F_F(t),A_F(t)]\cdots (F_F(t) + fI)). 
\end{eqnarray*} 
Hence we see that $kd\Psi_k - k(k-1)H\Psi_{k-1} = d/dt\tr((F_E(t) 
+ fI)^k - (F_F(t) + fI)^k)$.  Integrating 
from $t=0$ to $t=1$ shows that the $(1/(k-1)!)\int^1_0\Psi_k\ dt$ relate 
the two character forms $ch_H(\nabla_E,\nabla_F)$ and 
$ch_H(\nabla'_E,\nabla'_F)$.   

Recall that $\text{Mod}_{U_1}(L)$ denotes the 
semi-group of all $U_1$ bundle gerbe modules for 
the lifting bundle gerbe $L\to Y^{[2]}$ associated to 
a principal $PU(\mathcal{H})$ bundle $Y\to M$ with 
Dixmier-Douady class equal to $[H]$ (previously we 
were interested in $U_{\mathcal{K}}$ bundle gerbe 
modules but, as we have already mentioned, 
the extension of the theory to $U_1$ presents no 
difficulties).  
For a fixed bundle gerbe connection $\nabla_L$ on 
$L$ and a curving $f$ for $\nabla_L$, the  
character forms $ch_H(\nabla_E,\nabla_F)$ define a 
map $ch_H(L,\nabla_L,f)\colon \text{Mod}_{U_1}(L) 
\to H^\bullet(M;L,\nabla_L,f)$.  Specifically, $ch_H(L,\nabla_L,f)(E-F) 
= [ch_H(\nabla_E,\nabla_F)] \in H^\bullet(M;L,\nabla_L,f)$.  We have 
shown that this is independent of the choice of module 
connections $\nabla_E$ and $\nabla_F$ for $E$ and 
$F$.     
We need to investigate the effect that changing the 
bundle gerbe connection $\nabla_L$ on $L$ 
and the curving $f$ has on the character forms 
$ch_H(\nabla_E,\nabla_F)$.  It is only possible to 
change the bundle gerbe connection $\nabla_L$ on 
$L$ by $\delta(a)$ for some complex valued $1$-form 
$a$ on $Y$.  Then $\nabla_E$ and $\nabla_F$ no longer 
define module connections on $E$ and $F$, instead 
$\nabla_E - aI$ and $\nabla_F - aI$ define module 
connections on $E$ and $F$ for the new bundle gerbe 
connection $\nabla_L + \delta(a)$.  It is easy 
to check that $[ch_H(\nabla_E - aI,\nabla_F - aI)] = 
[ch_H(\nabla_E,\nabla_F)]$.  
Changing the curving $f$ by 
the pullback of a $2$-form $\lambda$ on $M$ to $Y$ 
changes the character forms by the exponential 
factor $\exp(\lambda)$; the maps $ch_H(L,\nabla_L,f)\colon 
\text{Mod}_{U_1}(L)\to H^\bullet(M;L,\nabla_L,f)$ and 
$ch_H(L,\nabla_L,f+\pi^*\lambda)\colon \text{Mod}_{U_1}(L)\to 
H^\bullet(M;L,\nabla_L,f+\pi^*\lambda)$ are related by 
$ch_H(L,\nabla_L,f+\pi^*\lambda) = \exp(\lambda)ch_H(L,\nabla_L,f)$.  
Any two principal $PU(\mathcal{H})$ bundles 
$Y$ and $X$ with Dixmier-Douady class $[H]$ are isomorphic, 
and this isomorphism extends to an isomorphism of 
the lifting bundle gerbes $L\to Y^{[2]}$ and $J\to X^{[2]}$ 
associated to $Y$ and $X$ respectively.   
There exist isomorphisms $\text{Mod}_{U_1}(L) = 
\text{Mod}_{U_1}(J)$ and we write $\text{Mod}_{U_1}(M,[H])$ 
for this isomorphism class of semi-groups.  It is 
clear that the maps $ch_H(L,\nabla_L,f_L)$ 
and $ch_H(J,\nabla_J,f_J)$ 
are compatible under the isomorphisms $\text{Mod}_{U_1}(L) 
= \text{Mod}_{U_1}(J)$ and hence descend to define a map 
$ch_{[H]}\colon \text{Mod}_{U_1}(M,[H])\to H^\bullet(M;[H])$.  
Under the isomorphism $\text{Mod}_{U_1}(M,[H]) = 
\tilde{K}^0(M;[H])$ of 
Proposition~\ref{prop:U_K bg modules equals K(M;H)} we 
get a map 
$$
ch_{[H]}\colon \tilde{K}^0(M;[H])\to 
H^\bullet(M;[H]).
$$
We propose that this map 
defines the Chern character for (reduced) twisted 
$K$-theory.  It can be shown that the 
Chern character for twisted $K$-theory 
is uniquely characterised by requiring that 
it is a functorial homomorphism which is compatible 
with the $\tilde{K}^0(M)$-module 
structure on $\tilde{K}^0(M;[H])$ and reduces to 
the ordinary Chern character when $[H] = 0$.  
It is easy to check that the map $ch_{[H]}
\colon \tilde{K}^0(M;[H])\to H^\bullet(M;[H])$   
is functorial with respect to smooth maps 
$f\colon N\to M$.  To show that $ch_{[H]}$ is a 
homomorphism, it is sufficient to show that the various  
maps $ch_H(L,\nabla_L,f)\colon \text{Mod}_{U_1}(L)\to 
H^\bullet(M;L,\nabla_L,f)$ are homomorphisms.  To see this,   
recall that the semi-group 
structure of $\text{Mod}_{U_1}(L)$ 
is defined via the direct sum of $U_1$ bundle gerbe 
modules (the direct sum $E\oplus F$ acquires a 
$U_1$ reduction rather than a $U_1\times U_1$ 
reduction from a fixed isomorphism 
$\mathcal{H}\oplus \mathcal{H} = \mathcal{H}$).  
>From here it is easy to see that $ch_H(\nabla_{E_1} \oplus 
\nabla_{E_2},\nabla_{F_1}\oplus \nabla_{F_2}) = 
ch_H(\nabla_{E_1},\nabla_{F_1}) 
+ ch_H(\nabla_{E_2},\nabla_{F_2})$.  We do not show here 
that $ch_{[H]}$ is compatible with the $\tilde{K}^0(M)$-module 
structure of $\tilde{K}^0(M;[H])$ (this can easily be 
shown to be true when $[H]$ is torsion, it is more 
difficult to prove this when $[H]$ is not torsion).

% -------------------------------------------------------------------------
\section{Conclusion}

Let us conclude with a final remark about $C^*$ algebras and bundle gerbes.
There is a well-known construction of a continuous
trace  $C^*$ algebra from
a groupoid \cite{Ren}.  This can be used to construct a $C^*$ algebra
from some bundle gerbes as follows.  If the fibres of
       $Y \to M$ have an appropriate measure on them then we can
define a product on two sections $f, g \colon  Y^{[2]} \to P$ by
$$
(fg)(y_1, y_2) = \int f(y_1, y)g(y, y_2) dy
$$
where in the integrand we use the bundle gerbe product so that
$f(y_1, y) g(y, y_2) \in L_{(y_1, y_2)}$.  Appropriately
closing this space of sections gives a $C^*$ algebra with spectrum $M$
and Dixmier-Douady class the Dixmier-Douady class of $(L, Y)$.
Some constructions in the theory of $C^*$ algebras become easy from this
perspective. For example if $A$ is an algebra with spectrum $X$ and
$f \colon Y \to X$
is a continuous map there is an algebra $f^{-1}(A)$ with spectrum $Y$.
This is just the pullback of bundle gerbes.  It is tempting to define
the $K$-theory of a bundle gerbe to be the $K$-theory of the
associated $C^*$ algebra.
However a result from \cite{Mur} is an obstruction to this. If a
bundle gerbe has non-torsion Dixmier-Douady class then the fibres of
$Y \to M$ are either infinite-dimensional (in which case there is no
measure) or disconnected. The simplest example of the disconnected
case is the one originally used by Raeburn and Taylor \cite{RaeTay},
in their proof that every three class is the Dixmier-Douady class of
some $C^*$ algebra, which is to take $Y$ the disjoint union of an
open cover so the fibres are discrete and counting measure
suffices. It follows from general theory however that when the $C^*$
algebra can be defined, because it has the same Dixmier-Douady class
as the bundle gerbe, its $K$-theory is the $K$-theory of the bundle
gerbe.

\bigskip \noindent
{\it Errata to \cite{BM}}: Bouwknegt and Mathai would like to correct
the following errors.\\
$\bullet$ page 5 in \cite{BM}, last paragraph. The $\mathcal
K$-bundles with torsion
Dixmier-Douady class $[H]$ described there are those that are pulled back
from the classifying space of the fundamental group, even though it is
not explicitly mentioned there. In general there are
$\mathcal K$-bundles with torsion
Dixmier-Douady class that can not be described in this manner.\\
$\bullet$ page 7 in \cite{BM}, in section 3. Since $\mathcal K$ has
no unit, one has
to add a unit in defining the $K$-theory of the algebra of sections
of the bundle of compact operators ${\mathcal E}_{[H]}$. The definition
given in page 7 works only in the case when the Dixmier-Douady class
$[H]$ is torsion, since in this case it can be shown that that the relevant
algebra of sections has
an approximate identity of idempotents, cf. \cite{Bla} \S 5.5.4.
This affects the discussion in the remaining part of the section
starting from the last paragraph on page 8, in
the sense that it is valid only when the Dixmier-Douady class $[H]$
is torsion. In particular, the finite dimensional description of
elements in twisted $K$-theory is valid only in the torsion case.
In the general case, sections of the twisted Fredholm
operators as in equation $(3.2)$ in \cite{BM} define elements in
twisted $K$-theory. \\
$\bullet$ page 8, equation $(3.3)$ in \cite{BM} should read
$$K^1(X, [H]) = [Y, \UK]^{Aut(\mathcal K)}$$
where $\UK$ is the group of unitaries on a
Hilbert space $\mathcal H$ of the form, identity operator
   + compact operator.
    
% -------------------------------------------------------------------------

% --------------------------------------------------------------------
\end{document}